\documentclass[amsmath,amssymb,
 aps, reprint, floatfix, superscriptaddress]{revtex4-2}
 
\pdfoutput=1
\usepackage[utf8]{inputenc}
\usepackage{graphicx}
\usepackage{dcolumn}
\usepackage{bm}
\usepackage{lipsum}
\usepackage{braket}
\usepackage{verbatim}
\usepackage{amssymb}
\usepackage{mathtools}
\usepackage{bbold}

\usepackage{csquotes}
\usepackage[T1]{fontenc}
\usepackage[english]{babel}
\usepackage{url}           
\usepackage{booktabs}      
\usepackage{amsfonts, amsmath, amssymb}      
\usepackage{nicefrac}       
\usepackage{microtype} 
\usepackage{float}
\usepackage{graphicx}
\usepackage{color}
\usepackage{subfigure}
\usepackage{hyperref}

\newcommand{\beq}{\begin{equation}}
\newcommand{\eeq}{\end{equation}}
\newcommand{\bea}{\begin{eqnarray}}
\newcommand{\eea}{\end{eqnarray}}
\newcommand{\nn}{\nonumber \\}
\newcommand{\sighat}{\hat{\sigma}}
\newcommand{\hhat}{\hat{H}}

\begin{document}

\title{The Perturbed Ferromagnetic Chain: A Tuneable Test of Hardness in the Transverse-Field Ising Model}

\author{D.T. O'Connor}
\affiliation{Department of Electronic and Electrical Engineering, University College London, Gower Street, London, UK}
\affiliation{London Centre for Nanotechnology, Gordon Street,
London, UK}
\author{L. Fry-Bouriaux}
\affiliation{London Centre for Nanotechnology, Gordon Street,
London, UK}
\author{P.A. Warburton}
\affiliation{Department of Electronic and Electrical Engineering, University College London, Gower Street, London, UK}
\affiliation{London Centre for Nanotechnology, Gordon Street,
London, UK}
\date{\today}

\begin{abstract}
    Quantum annealing in the transverse-field Ising model (TFIM) with open-system dynamics is known to use thermally-assisted tunneling to drive computation. However, it is still subject to debate whether quantum systems in the presence of decoherence are more useful than those using classical dynamics to drive computation. We contribute to this debate by introducing the perturbed ferromagnetic chain (PFC), a chain of frustrated sub-systems where the degree of frustration scales inversely with the perturbation introduced by a tunable parameter. This gives us an easily embeddable gadget whereby problem hardness can be tuned for systems of constant size. We outline the properties of the PFC and compare classical spin-vector Monte Carlo (SVMC) variants with the adiabatic quantum master equation. We demonstrate that SVMC methods get trapped in the exponentially large first-excited-state manifold when solving this frustrated problem, whereas evolution using quantum dynamics remains in the lowest energy eigenstates. This results in significant differences in ground state probability when using either classical or quantum annealing dynamics in the TFIM.  
\end{abstract}

\maketitle

\section{Introduction}
Experimental validation of quantum processes and/or computational scaling advantages in adiabatic quantum computation typically relies on the use of artificial gadgets \cite{nagajQuantumSpeedupQuantum2012,denchevWhatComputationalValue2016, boixoComputationalMultiqubitTunnelling2016, albashDemonstrationScalingAdvantage2018, mandraDeceptiveStepQuantum2018, kingScalingAdvantagePathintegral2021}, as these provide a way to demonstrate quantum dynamics that can be exploited to enhance computation. A possible framework to implement these gadgets is through the transverse-field Ising model (TFIM), where the quantum dynamics are introduced through the addition of a local non-commuting Hamiltonian to the system. If one interpolates between the non-commuting Hamiltonians such that the system ends in the computational basis of a problem space, then this is known as quantum annealing \cite{kadowakiQuantumAnnealingTransverse1998, farhiQuantumAdiabaticEvolution2001}. Such Hamiltonians that use this method typically have the form
\beq\label{eq. exp_Ham}
\hat{H}(s) = -A(s) \sum_{i=1}^{N}\sighat_{i}^{x} + B(s) \left[\sum_{i=1}^{N}h_{i}\sighat_{i}^{z} + \sum_{i,j}J_{ij}\sighat_{i}^{z}\sighat_{j}^{z} \right] \,,
\eeq
where $\sighat^{x}$ and $\sighat^{z}$ are the Pauli X and Z matrices respectively, and the $n$-qubit problem is encoded in the biases, $h_i$, and couplers, $J_{ij}$. The coefficients $A(s)$ and $B(s)$ are positive monotonically decreasing and increasing functions in normalized time, $s$, respectively. Typically the system is interpolated from the first term, which is the transverse field component that has an easy-to-find ground state solution at $s = 0$, to the second term that encodes the problem of interest at $s = 1$.

This Hamiltonian has been extensively studied on experimental quantum annealers for the past two decades, both within the context of combinatorial optimization \cite{venturelliReverseQuantumAnnealing2019, stollenwerkFlightGateAssignment2018, kimLeveragingQuantumAnnealing2019, inoueTrafficSignalOptimization2021, kitaiDesigningMetamaterialsQuantum2020, asproniAccuracyMinorEmbedding2020, neukartTrafficFlowOptimization2017} and quantum simulation \cite{kingObservationTopologicalPhenomena2018, kingScalingAdvantagePathintegral2021, bandoProbingUniversalityTopological2020, harrisProbingNoiseFlux2008, kairysSimulatingShastrySutherlandIsing2020}. However, the quantumness of the dynamics used for computation on quantum annealers is still subject to debate due to the prevalent quantum and classical noise sources that can obscure coherent quantum processes~\cite{crowleyQuantumClassicalDynamics2014, albashDecoherenceAdiabaticQuantum2015, zaborniakBenchmarkingHamiltonianNoise2021, lantingProbingHighfrequencyNoise2011}. 

One such process is that of thermalization, which has been seen to aid computation on quantum annealers where the anneal times are orders of magnitude larger than the single-qubit decoherence time~\cite{dicksonThermallyAssistedQuantum2013}. Improvements to ground state probability can also be realised if we pause mid-anneal and allow the system to thermalize near the minimum gap~\cite{marshallPowerPausingAdvancing2019, albashComparingRelaxationMechanisms2021, chenWhyWhenPausing2020}. However, to what extent thermalization is beneficial is still an active area of research, with lower noise quantum annealers (i.e. those with lower thermalization rates) being seen to improve tunneling ranges in local search (reverse) quantum anneals~\cite{chancellorExperimentalTestSearch2021}. It was also shown in Ref.~\cite{albashComparingRelaxationMechanisms2021} that the thermalization signature of a quantum anneal can be replicated with spin-vector Monte Carlo (SVMC) \cite{shinHowQuantumDWave2014}, which is a classical heuristic used to mimic the behaviour of physical quantum annealers~\cite{muthukrishnanTunnelingSpeedupQuantum2016, albashDemonstrationScalingAdvantage2018}. Therefore it is important to ask where thermalization with quantum dynamics can be useful computationally when annealing in the TFIM.

In order to help answer this, we introduce the perturbed ferromagnetic chain (PFC), a scaleable and tunable gadget that we use to differentiate between SVMC and a system annealed under quantum dynamics. The PFC possesses the qualities of a having a false minimum during an anneal (if the magnitude of perturbation is small enough) and has an exponentially large degenerate first excited state manifold in the computational basis. Additionally we can tune the minimum gap energy, $\Delta_{10}$, with the perturbative parameter such that it can be tuned through the value of the environmental temperature. 

In the thermal regime, it is possible to compare the computational use of thermally assisted quantum tunneling with classical mechanisms. The extent of the former is analytically explored using the quantum adiabatic master equation (AME) \cite{albashQuantumAdiabaticMarkovian2012, albashDecoherenceAdiabaticQuantum2015}, whereby we simulate quantum annealing using open system dynamics. This provides a model of tunneling in a system that experiences decoherence at a finite temperature, such that we can observe its computational role when annealing with the PFC.

The format of the paper is as follows. We outline the definition of the PFC in section~\ref{section_pfc} and give an overview of its properties, both classically at thermal equilibrium and in the TFIM by inspecting the behaviour of both the quantum and semi-classically approximated states. This will set the foundation as to why this is a problem of interest when testing the quantumness of certain methods in the TFIM. In section~\ref{section_methods} we introduce the methods for simulating both the quantum system using the AME and the classical system using SVMC. In order to fully explore the dynamics of SVMC, we look at both SVMC and SVMC-TF (SVMC with transverse-field dependent updates~\cite{albashComparingRelaxationMechanisms2021}) as well as introducing an additional degree of freedom into both variants, which will allow for full exploration of the Bloch sphere. Finally in section~\ref{section_results} we present the dynamical simulation results and demonstrate the effect that the false minimum and the exponentially large first excited state manifold has on ground state probability for both the AME and SVMC. We look at these effects with respect to both the magnitude of perturbation and the PFC system size.

\section{Perturbed Ferromagnetic Chain}
\label{section_pfc}

\subsection{Classical Model}
\label{Methods_problem}

\begin{figure}
\includegraphics[width=\columnwidth]{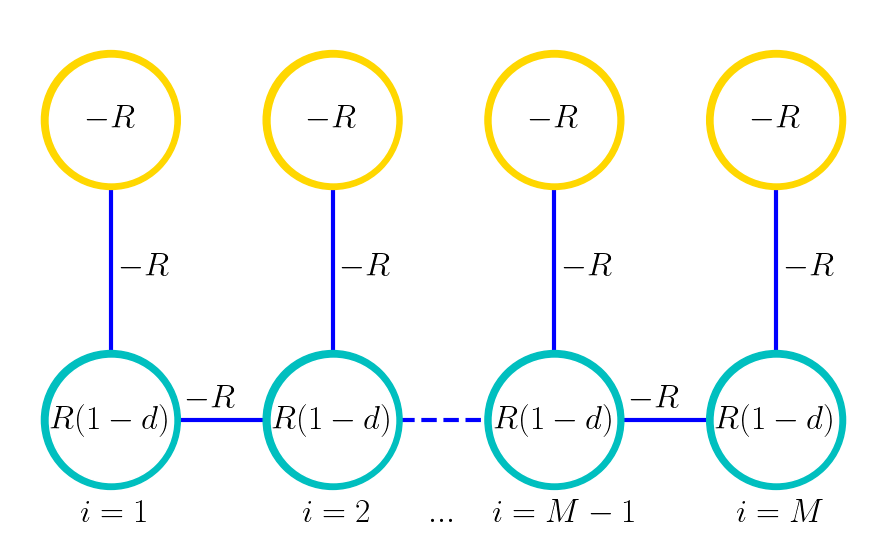}
\caption{Illustration of the PFC Hamiltonian in general form. In yellow are the auxiliary qubits with biases of $-R$, and in turquoise are the backbone qubits with biases $R(1-d)$. The dark blue edges are the ferromagnetic couplers of strength $-R$. The total number of qubits, $N$, in this problem is $2M$, where $M$ denotes the number of 2-qubit sub-systems (indexed by $i$) in the Hamiltonian. The system energy is scaled by $R$, and $d$ describes the magnitude of perturbation. The properties of this model hold generally for $M \geq 2$, $R > 0$ and $1 > d > 0$.}
\label{fig:gadget}
\end{figure}

The PFC (Fig.~\ref{fig:gadget}) is a ferromagnetically coupled chain of frustrated sub-systems each composed of two qubits. This system is similar to the cyclic spin-gadgets used in Refs.~\cite{boixoExperimentalSignatureProgrammable2013, albashConsistencyTestsClassical2015}, but instead the gadget is made to be acyclic and a perturbative offset, $d$, is introduced to break the degeneracy of the ground state (at $d = 0$) into a single ground state and $2^M$-degenerate first excited state. The degree of frustration in the PFC scales inversely with $d$. The PFC Hamiltonian is given by
\bea
&  \frac{1}{R} \hhat_P =  \hhat_{ss} - \sum^{M-1}_{i=1}\sighat^z_{b, i}\sighat^z_{b, i+1} \,, \label{eq:prob_ham}& \\
&  \hhat_{ss} = \sum^M_{i=1} (1-d)\sighat^z_{b, i} - \sighat^z_{a, i} - \sighat^z_{a, i}\sighat^z_{b, i} \,, \nonumber & 
\eea
\noindent where $M$ is the number of subsystems, $R$ scales the energy of the problem, and the magnitude of the perturbation is characterized by the parameter $d$. The auxiliary and backbone qubits, depicted by the yellow and turquoise circles shown in Fig.~\ref{fig:gadget} respectively, are denoted by Pauli Z matrices $\sighat^z_{a}$ and $\sighat^z_{b}$ respectively. The ground state of this Hamiltonian is the $|0_b^{\otimes M}, 0_a^{\otimes M} \rangle$ (all up) state, and given that $d < (M-1)^{-1}$ the first excited state is a degenerate manifold whose size grows exponentially in $M$. This manifold always has the backbone qubits in the $|1_b \rangle^{\otimes M}$ (all down) configuration, and the auxiliary qubits are iso-energetic with respect to their spin state, creating a $2^M$-degenerate manifold of "floppy" auxiliary qubits. This creates an energy gap between the ground state and exponential manifold of $\Delta_{10} = 2RMd$, where the states in the manifold have a Hamming distance between $M$ and $2M$ from the ground state. 

Classically the PFC is exactly solvable via the transfer matrix method \cite{kramersStatisticsTwoDimensionalFerromagnet1941}, where at an inverse temperature, $\beta$, the partition function, $\mathcal{Z}$, can be found in polynomial time. The partition function in transfer matrix form is expressed as
\bea
&
\mathcal{Z} = \mathbf{v} \mathbf{W}^{M-1} \mathbf{v}^T
\, , \label{eq:classical_Z}
\eea
where,
\bea
\mathbf{v} = 
\begin{pmatrix}
e^{\frac{1}{2} \beta R \left( d+1 \right)} & e^{\frac{1}{2} \beta R \left( d-3 \right)} & e^{\frac{1}{2} \beta R \left( 1-d \right)} & e^{\frac{1}{2} \beta R \left( 1-d \right)} \\ 
\end{pmatrix}
\label{eq: boundary_vec}
\eea
handles the boundary subsystems of the chain, and
\bea
\mathbf{W} = 
\begin{bmatrix} 
e^{\beta R\left( d+2 \right)} & e^{\beta R d } & 
1 & 1 \\
e^{\beta R d} & e^{\beta R \left( d - 2 \right)} & 
e^{-2\beta R} & e^{-2\beta R} \\
1 & e^{-2\beta R} & 
e^{\beta R \left( 2 - d \right)} & e^{\beta R \left( 2-d \right)} \\
1 & e^{-2\beta R} & 
e^{\beta R \left( 2-d \right)} & e^{\beta R \left( 2-d \right)}
\end{bmatrix} ,
\label{eq: transfer_matrix}
\eea
handles the inner subsystems of the chain. We can then find the magnetization of a subsystem at thermal equilibrium using
\bea
\langle \sigma^z_i \rangle = \frac{1}{\mathcal{Z}} \left[ \mathbf{v} \mathbf{W}^{i-1} \sigma^z_i \mathbf{W}^{M-i} \mathbf{v}^T \right]  \,, \label{eq:transfer_mag}
\eea
where $\sigma^z_i = \frac{1}{2}\left(\sighat^z_{a,i} + \sighat^z_{b,i}\right)$. The average magnetization of the PFC is then an average over all contributions, $\langle \sigma^z \rangle = \frac{1}{M}\sum_i^M \langle \sigma^z_i \rangle$.

Additionally, the free-energy of the PFC can be derived from the transfer matrix, which can then be used to derive further thermodynamic properties. The free-energy, $F$, is defined as
\beq
F = -\lim_{M \rightarrow \infty} \frac{1}{\beta M} \ln \mathcal{Z}\,.
\label{eq: free energy def}
\eeq
We show in Appendix~\ref{Appendix_free_energy} that in the limit of $M \rightarrow \infty$, that $\mathcal{Z}^{1/M} \rightarrow \lambda_1$, where $\lambda_1$ is the spectral radius (largest absolute eigenvalue) of $\mathbf{W}$. The free-energy of the PFC is found to be
\begin{equation}
\begin{split}
    F & =  - \frac{1}{\beta} \ln \Bigg( e^{2\beta R} \cosh{\beta Rd} + \cosh{\beta R(2-d)} +\\ & \sqrt{\left( e^{2\beta R} \cosh{\beta Rd} + \cosh{\beta R(2-d)} \right)^2 - 4\sinh{4\beta R}} \Bigg)\,,
\end{split}
    \label{eq: free_energy_PFC}
\end{equation}
which is always real, continuous and finite given that the original parameter constraints of the PFC are met.

\subsection{Transverse-Field Ising Model}
\label{quantum_pfc}

\begin{figure*}[ht!]
    \subfigure[\,]{\includegraphics[width=\columnwidth]{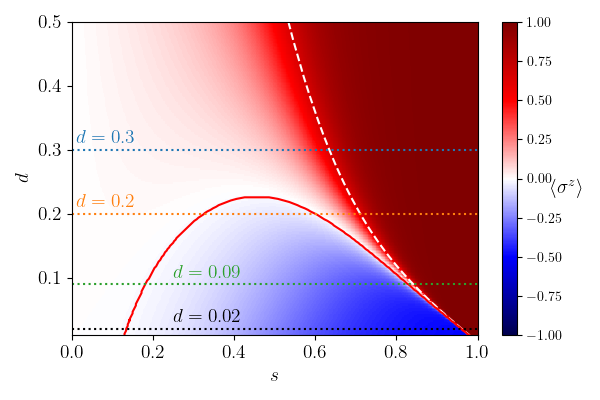}
    \label{fig:instant_mag_e0}}
    \subfigure[\,]{\includegraphics[width=\columnwidth]{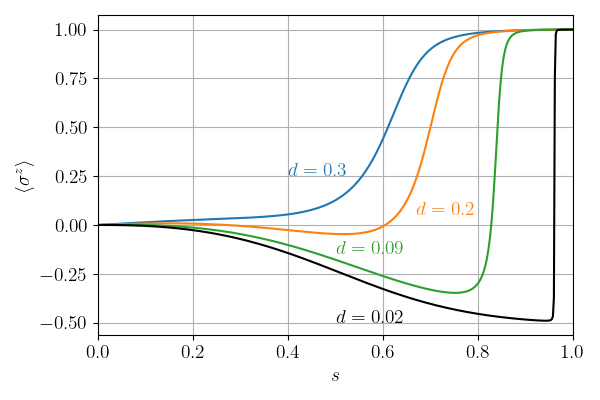}
    \label{fig:instant_mmag_e1}}
    \caption{(a) Intensity plot of the average qubit magnetization in the instantaneous ground state for a PFC in the presence of a transverse-field (Eq.~(\ref{eq:prob_ham_quantum})) with $M=2$ and $R=1.0$. The red line shows the boundary between the quantum paramagnetic and negative magnetization phases. The white dashed line indicates the position of the minimum gap. (b) Cross-sections of (a) showing the average magnetization during an anneal.}
    \label{fig:magnetization}
\end{figure*}

\begin{figure*}[ht!]
    \includegraphics[width=2\columnwidth]{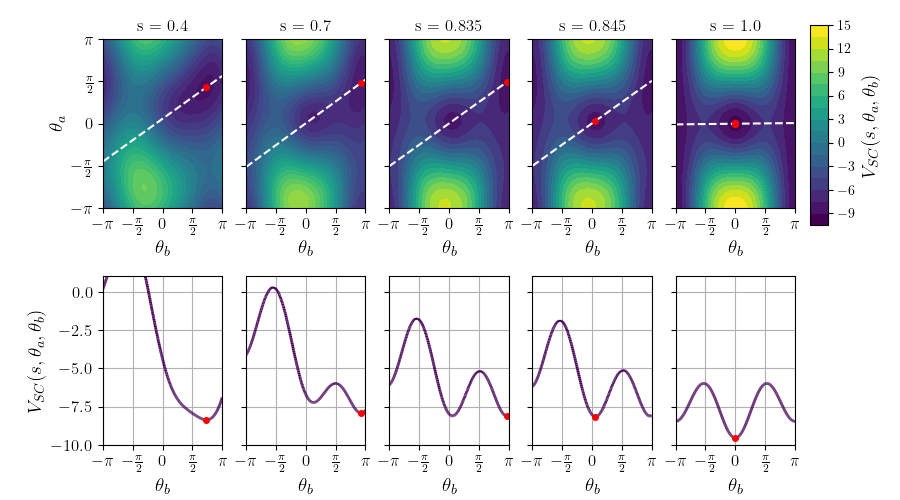}
    \caption{Plots of the semi-classical energy potential (top row), and the energy along the hyper-plane (white dashed line) passing through the landscape (bottom row) at the specified values of normalized time, $s$. This potential is for a PFC with $M=2$, $R=1.0$ and $d = 0.09$. The red marker indicates the global minimum of the landscapes. The backbone and auxiliary spins in the PFC are parameterized by angles $\theta_b$ and $\theta_a$ respectively.}
    \label{fig:semi-classical}
\end{figure*}

Translating the PFC into the TFIM involves the addition of a non-commuting transverse field term, composed of $\sighat^x$ operators, which introduces the driver of quantum fluctuations that can potentially be used to aid computation~\cite{kadowakiQuantumAnnealingTransverse1998, farhiQuantumAdiabaticEvolution2001}. The TFIM Hamiltonian of the PFC is given by  
\beq\label{eq:prob_ham_quantum}
\hat{H}(s) = -A(s) \sum_{j=1}^{N}\sighat_{j}^{x} + B(s) \hhat_P \,,
\eeq
where the classical PFC is encoded into $\hhat_P$ (Eq.~(\ref{eq:prob_ham})). The coefficients are taken to be $A(s) = 3(1-s)$ and $B(s) = 3s$ throughout this work, where $s$ is the normalised annealing time $s = t / t_{\textrm{anneal}}$. 


For sufficiently small values of $d$, the ground state of this Hamiltonian is seen to undergo a quantum phase transition, illustrated in Fig.~\ref{fig:magnetization} by the change in average qubit magnetization from negative to positive phases. The average qubit magnetization is defined as 
\beq
\langle \sigma^z \rangle = \frac{1}{N} \sum_{j=1}^{N} \langle E_0 (s) | \sighat^z_j | E_0 (s) \rangle \,,
\eeq
where $| E_0 (s) \rangle$ is the instantaneous ground state from the diagonalized Hamiltonian of Eq.~(\ref{eq:prob_ham_quantum}) at some value of $s$, and $N$ is the number of qubits in the PFC. The formation of the negative phase before the minimum gap is indicative of the ground state qubits becoming magnetized to resemble the exponentially large degenerate first-excited-state manifold (further illustrated in sections~\ref{Methods_semiclassical} and~\ref{section_results}). After passing through the minimum gap, the instantaneous ground state enters the positive phase and then goes on to finish in the computational ground state. It must be noted that even if the $d < (M-1)^{-1}$ condition is broken such that the exponentially degenerate manifold is no longer the computational first-excited-state, for small $d$ the instantaneous ground state maintains its resemblance to the exponentially degenerate manifold before the minimum gap.  

The tuning of $d$ scales the size of the gap for all $M$ in the TFIM, and therefore can be used as a tunable hardness parameter for the PFC. Although scaling with $M$ is numerically seen to exponentially scale the minimum gap size and therefore hardness, it can lead to intractable computational times for some of the simulation methods explored in the results, making $d$ a more desirable tunable hardness parameter. 

In summary, the PFC becomes hard in the TFIM for small values of $d$ due to the presence of a quantum phase transition, and for large values of $M$ where the minimum gap exponentially reduces in size. In the next section we perform a semi-classical analysis to show that when the PFC is translated into to the TFIM a false minimum exists. The interplay between the false and true minima is particularly prominent when annealing through the region where the ground state is in the negative magnetization phase shown in Fig.~\ref{fig:magnetization}.

\subsection{Semi-Classical Analysis}
\label{Methods_semiclassical}

To further explore the behaviour of the PFC in the TFIM, a semi-classical approximation can be made by using the spin-coherent Ansatz~\cite{klauderPathIntegralsStationaryphase1979}
\beq 
  \left| \theta \right\rangle = \bigotimes^N_{j=1} \cos\left(\frac{\theta_j}{2}\right) |0\rangle + e^{i\phi_j} \sin\left(\frac{\theta_j}{2}\right) |1\rangle
  \label{eq:sc_ansatz_full}
\eeq
to calculate the semi-classical effective potential landscape as a function of $s$. The magnetization expectation values of the auxiliary qubits are almost identical, and the same is true for the backbone qubits. We can therefore approximate the states of the PFC in the spin-coherent Ansatz as
\begin{align}
    \left| \theta_a, \theta_b \right\rangle &=
    \begin{aligned}
         & \left[\bigotimes^M_{i=1} \cos\left(\frac{\theta_a}{2}\right) |0\rangle + \sin\left(\frac{\theta_a}{2}\right) |1\rangle \right] \otimes &  \\
        & \left[\bigotimes^M_{i=1} \cos\left(\frac{\theta_b}{2}\right) |0\rangle + \sin\left(\frac{\theta_b}{2}\right) |1\rangle \right] \,.
    \end{aligned}
    \label{eq: sc_ansatz} 
\end{align}
Here, $\theta_a$ and $\theta_b$ are the angles of the states in the XZ-plane of the Bloch spheres for all of the auxiliary and backbone qubits respectively. Here we assume that the azimuthal angle, $\phi_j$ is equal to zero. The semi-classical potential is then given by
\begin{equation}
     V_{SC}(s, \theta_a, \theta_b) = \left \langle \theta_a, \theta_b \right| \hat{H}(s) \left|\theta_a, \theta_b \right\rangle.
     \label{eq: sc_potential} 
\end{equation}

The visual representation of the potential at various stages of an anneal (Fig.~\ref{fig:semi-classical}) shows the PFC initially taking a path to the first excited states ($\theta_a \in [-\pi, \pi], \theta_b = \pi$). This is then followed by a discontinuous change in the position of the global energy minimum about the minimum gap (at $s = 0.841$) to the computational ground state ($\theta_a = 0, \theta_b = 0$). By taking a hyper-plane that passes through the global minimum (and the local minimum where applicable), it is clear that as we evolve from a uni-modal to a bi-modal potential the all down ($\theta_b = \pi$) configuration of the backbone qubits is energetically preferable until the minimum gap is traversed. 

Additionally, the computational ground state is energetically isolated from the low-energy excited states, meaning that further dynamical evolution is still needed to reach the ground state after the minimum gap. If the system is evolved under an adiabatic, coherent regime~\cite{katoAdiabaticTheoremQuantum1950} then the dynamical process is quantum tunneling. In a classical model (like SVMC) we can only use thermal excitations to traverse these energy barriers. 

\begin{figure}
    \includegraphics[width=\columnwidth]{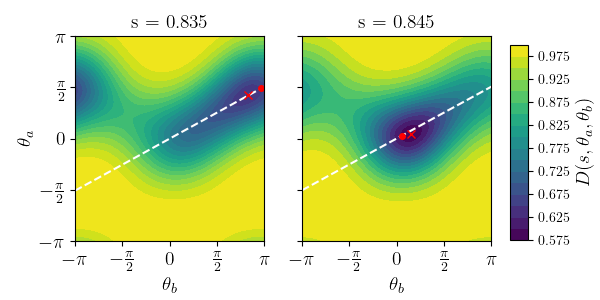}
    \caption{The trace-norm distance between the spin-coherent state and the instantaneous ground state (Eq.~(\ref{eq:trace-norm})) in the vicinity of the minimum gap at $s = 0.841$ for a PFC with $M=2$, $R=1.0$ and $d = 0.09$. The backbone and auxiliary spins in the PFC are characterised by angles $\theta_b$ and $\theta_a$ respectively. The red cross marker indicates the global minimum of the trace-norm landscapes and the red circle indicates the minimum of the potential landscape in Fig.~\ref{fig:semi-classical}. The hyper-plane (white dashed line) from Fig.~\ref{fig:semi-classical} is also plotted.}
    \label{fig:semi-class_norm}
\end{figure}

If tunneling were to occur in the instantaneous ground state, then this would result in delocalization about the bistable potential. Using the trace-norm distance,
\beq
D(s, \theta_a, \theta_b) = \sqrt{1 - \left|\langle E_0 (s) | \theta_a, \theta_b \rangle \right|^2 } \,,
\label{eq:trace-norm}
\eeq
we can quantify the distance between the instantaneous ground state and the spin-coherent Ansatz, to determine the extent to which the Ansatz accurately describes the instantaneous ground state. In Fig.~\ref{fig:semi-class_norm} we show the trace-norm distance for a four-qubit ($M=2$) instance of the PFC in the vicinity of the minimum gap. This is the same instance whose potential landscape is shown in Fig.~\ref{fig:semi-classical} and whose minimum gap occurs at $s = 0.841$. At $s = 0.835$ the trace-norm distance has a global minimum whose location in $( \theta_a , \theta_b )$ space closely corresponds to the global minimum of the semi-classical potential (i.e. near $ \theta_b = \pi$). Nevertheless there is a local minimum in the trace-norm distance which extends along the indicated hyper-plane, showing that, prior to the minimum gap, tunneling enables a finite probability amplitude in the local minimum of the potential landscape near $ \theta_b = 0$. After the minimum gap, as shown for the case $s = 0.845$ in the right panel of Fig.~\ref{fig:semi-class_norm}, the global minimum of the trace-norm distance now closely corresponds to the global minimum of the semi-classical potential near $ \theta_b = 0$. This minimum continuously evolves into the global minimum of the problem Hamiltonian at $ s = 1$ as shown in Fig.~\ref{fig:semi-classical}. 

Using semi-classical analysis we have shown the existence of a false global minimum before the minimum gap, when the value of $d$ is small enough. Under quantum evolution, the transition from the false minimum to the true minimum exploits tunnelling, and this is visualized by measuring the trace-norm distance between spin-coherent states and the instantaneous ground state to show delocalization across the potential barrier (Fig.~\ref{fig:semi-class_norm}). Under classical evolution, thermal excitation of multiple qubits is needed to traverse the barrier to reach the computational ground state. Additionally the manifold along $\theta_b = \pi$ (corresponding to the exponentially large computational first excited state manifold) is equally accessible under classical dynamics. Therefore classical algorithms that explore this energy landscape, such as SVMC, can remain in this manifold instead of reaching the ground state. We explore in section~\ref{section_results} the extent to which this hinders computation in both classical and quantum evolution. 
 
\section{Methods}
\label{section_methods}

\subsection{Spin-Vector Monte Carlo}
\label{Methods_SVMC}

The spin-vector Monte Carlo algorithm \cite{shinHowQuantumDWave2014} is an algorithm that replaces the Pauli matrices in the Hamiltonian of Eq.~(\ref{eq:prob_ham_quantum}) with $O(2)$ rotors in the XZ-plane of the Bloch-sphere. The energy function we are to minimize using Metropolis-Hastings updates becomes
\beq  \label{eq:SVMC_energy_fn}
\begin{split}
\mathcal{E}(s) & = -A(s)\sum^N_{j=1} \sin \theta_j +  \\ 
& B(s)\left[\sum^N_{j=1} h_j \cos \theta_j + \sum_{\langle j, k \rangle} J_{jk} \cos \theta_j \cos \theta_k \right] \,,
\end{split}
\eeq
where $A(s)$ and $B(s)$ are the same schedule functions introduced in Eq.~(\ref{eq. exp_Ham}). The SVMC algorithm attempts to update all rotor angles, $\theta$, in every sweep (i.e. every increment of time). The method of update can be described in two ways: 
\bea
    \label{svmc_update}
    & \theta_j^t \in [0, \pi], \quad \theta_j^{t+1} \sim \textrm{Uniform}(0, \pi) \,,  & \\ 
    & \theta_j^t \in [0, \pi], \quad \theta_j^{t+1} = \theta_j^t + \min\left( \frac{A(s)}{B(s)}, 1 \right) u, & \nn
    \label{svmctf_update}
    & \quad u \sim \textrm{Uniform}(-\pi, \pi) \,. &
\eea
The traditional method of update in SVMC is described by Eq.~(\ref{svmc_update}), whereby the new angle is a sample from a uniform distribution from zero to $\pi$. A more recent update method used to capture additional annealing artifacts such as freeze-out is shown in Eq.~(\ref{svmctf_update}), which we refer to as SVMC-TF  \cite{albashComparingRelaxationMechanisms2021}. In this latter version, the freedom of the rotor movement in an update is proportional to the relative magnitude of the transverse field that drives the dynamics.

We also consider another variant of SVMC and SVMC-TF, whereby the dynamical restriction of only operating in the XZ-plane is removed by including the azimuthal angle $\phi_j$ in Eq.~(\ref{eq: sc_ansatz}), such that SVMC now has access to the entire Bloch sphere. We will refer to these variants as spherical-SVMC and spherical-SVMC-TF. This coordinate extension does not affect the Z (polar) components of the energy function. It does however affect the transverse field component, such that the new energy function becomes

\beq  \label{eq:SVMC_sph_energy_fn}
\begin{split}
\mathcal{E}(s) & = -A(s)\sum_{j=1}^N \cos \phi_j \sin \theta_j +  \\ 
& B(s)\left[\sum_{j=1}^N h_j \cos \theta_j + \sum_{\langle j, k \rangle} J_{jk} \cos \theta_j \cos \theta_k \right] \,,
\end{split}
\eeq
where the azimuthal angle, $\phi_j \in [-\pi, \pi]$, is also updated in the same way as the polar angle, $\theta_j$.

Throughout the rest of this paper, we look at both the simplest case, SVMC, and the more complex spherical-SVMC-TF for the comparative experiments, with the other variants included in appendix~\ref{Appendix_svmc_var} for completeness. For all variants, we also take the annealing functions to be $A(s) = 3(1-s)$ and $B(s) = 3s$, where we start the annealing from $s = 0$ to $s = 1$ at a temperature of $12$ mK. All algorithms update spins individually in a randomly permuted order, thus cannot capture any simultaneous multi-qubit moves unlike those that may occur in a system evolved using quantum dynamics.

\subsection{Quantum Dynamics}
\label{Methods_analytical}

To simulate the quantum evolution of the PFC, we look at the dynamics in both closed and open systems using the von-Neumann equation and adiabatic master equation~\cite{albashQuantumAdiabaticMarkovian2012, albashDecoherenceAdiabaticQuantum2015} (AME) respectively. For the closed system simulations, the von-Neumann equation takes the form, 
\beq
    \hbar \frac{\partial}{\partial t}\hat{\rho}(t) = -i \left[ \hat{H}(t), \hat{\rho}(t) \right] \,,
    \label{eq:von-neumann}
\eeq
where $\hat{\rho}(t)$ is the density matrix, and $t$ is related to normalized time, $s$, by $s = t/t_{\textrm{anneal}}$. The initial state is the pure ground state of the system at $s=0$, which when using Eq.~(\ref{eq:prob_ham_quantum}) is $\hat{\rho}(0) = |+\rangle \langle +|$. To solve the von-Neumann equation and the AME we have used the Hamiltonian Open Quantum System Toolkit~\cite{chenHOQSTHamiltonianOpen2020} (HOQST), a simulation library written for the Julia language.

The form of the AME we have chosen for the open system simulations is one that uses parameters similar to those used in theoretical studies of the experimental annealing hardware \cite{albashConsistencyTestsClassical2015, mishraFiniteTemperatureQuantum2018}. We will not define the AME in its entirety here (see Refs.~\cite{chenHOQSTHamiltonianOpen2020, albashQuantumAdiabaticMarkovian2012}), but we use a Davies-style AME that is valid in the weak-coupling limit. It is also necessary to make assumptions on the model of decoherence used, where we assume that all qubits are coupled equally to the bath independently. All qubits experience decoherence by dephasing, and the bath spectral density takes the form of a Bosonic Ohmic bath,
\begin{equation}
    \gamma(\omega) = 2 \pi \eta g^2 \frac{\omega \exp\left(-|\omega|/\omega_c\right)}{1 - \exp\left(-\beta \omega\right)} \,,
    \label{eq:bath_fn}
\end{equation}
\noindent where $\beta = (k_B T)^{-1}$ is the inverse temperature, $\omega_c$ is the cutoff frequency, $\eta g^2$ is the dimensionless bath coupling strength and $k_B$ is the Boltzmann constant. Throughout the rest of the paper we specify the bath parameters to be $T=12$ mK, $\omega_c = 4$ GHz and $\eta g^2 = 10^{-3}$. 

The decoherence by dephasing then manifests itself through the time-dependent Lindblad operators
\begin{equation}
    \hat{L}_{j, \omega_{kl}} (t) = \langle E_l (t) | \sigma^z_j | E_k (t) \rangle | E_l (t) \rangle \langle E_k (t) | \,.
    \label{eq:lindblad}
\end{equation}
This describes how the $j^{th}$ qubit couples to environment according to energy gaps, $\omega_{kl} = E_k - E_l$, between the instantaneous energy eigenstates, $| E_k (t) \rangle$ and $| E_l (t) \rangle$ of the system Hamiltonian (Eq.~(\ref{eq:prob_ham_quantum})).

We use this form of the AME since it models thermally assisted adiabatic quantum computation near the adiabatic limit. Assuming that most of the ground state population is lost to the first excited state after passing through the minimum gap, the re-population of the ground state via thermal relaxation can be related to the transition rate via
\begin{equation}
    \Gamma_{1 \rightarrow 0} (t) \propto \gamma_{1 \rightarrow 0} (t) = \gamma(\omega_{10}(t)) \sum_j \left| \langle  E_0 (t) | \sigma^z_j| E_1 (t) \rangle \right|^2 \,.
    \label{eq:transition}
\end{equation}
Here the temperature dependence of the transition rate is introduced by the Ohmic spectral density function, $\gamma$ in Eq.~(\ref{eq:bath_fn}). 

However, when the gap, $\omega_{10}$, is sufficiently small, the weak-coupling assumption in the AME starts to break down. This is the case for hard PFC instances since the gap can be very small relative to the bath temperature. In such a regime the energy levels become broadened due to the stronger coupling to the bath, such that the discrete energy levels should emulate a more continuous potential, similar to the semi-classical picture. Therefore, despite the AME not being able to describe these strong-coupling regimes as accurately as more sophisticated models like the Redfield equation, it serves as a reasonable approximation of an open-system model of the PFC.

\section{Dynamical Simulations}
\label{section_results}

\begin{figure}
    \centering
    \includegraphics[width=\columnwidth]{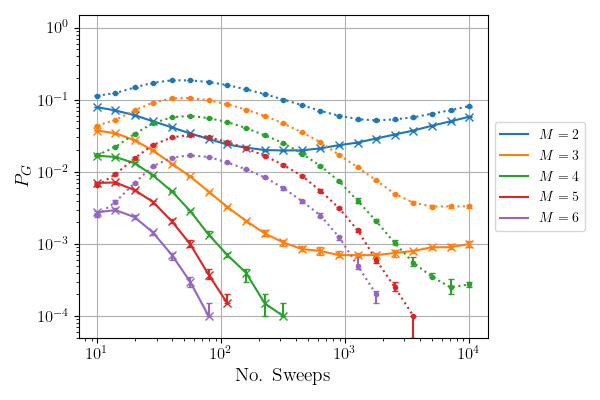}
    \caption{Probability of being in the ground state for both SVMC (solid lines) and spherical-SVMC-TF (dotted lines) as the system scales in size, $M$, for a PFC with $d = 0.1$. Here, the SVMC and spherical-SVMC-TF probabilities are found from 20,000 samples, which we repeat 50 times and bootstrap to find the median and $95\%$ confidence intervals for the data point and error bars respectively.}
    \label{fig:result_svmc_n_test}
\end{figure}
\begin{figure}
    \centering
    \includegraphics[width=\columnwidth]{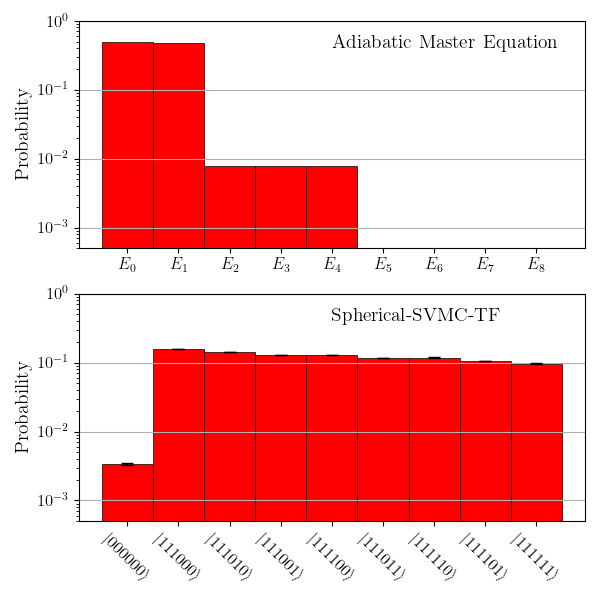}
    \caption{State Probability at $s = 0.83$ ($= t / t_{\textrm{anneal}}$), just after passing through the minimum gap ($s = 0.8227$) for a) the AME with $ t_{\textrm{anneal}} = 200 ns$ and b) spherical-SVMC-TF with $ t_{\textrm{anneal}} = 10,000$ sweeps. We measure a PFC with $M = 3$ and $d = 0.1$. For the AME we measure the probability of being in the $j^{th}$ instantaneous state, $E_{j}$, and for spherical-SVMC-TF we take a classical measurement of being in either the ground state or any of the first excited states. The spherical-SVMC-TF probabilities are found from 20,000 samples, of which we repeat 50 times and bootstrap to find a median and $95\%$ confidence intervals for the data point and error bars respectively.}
    \label{fig:result_populations}
\end{figure}

\begin{figure*}
    \centering
    \includegraphics[width=2\columnwidth]{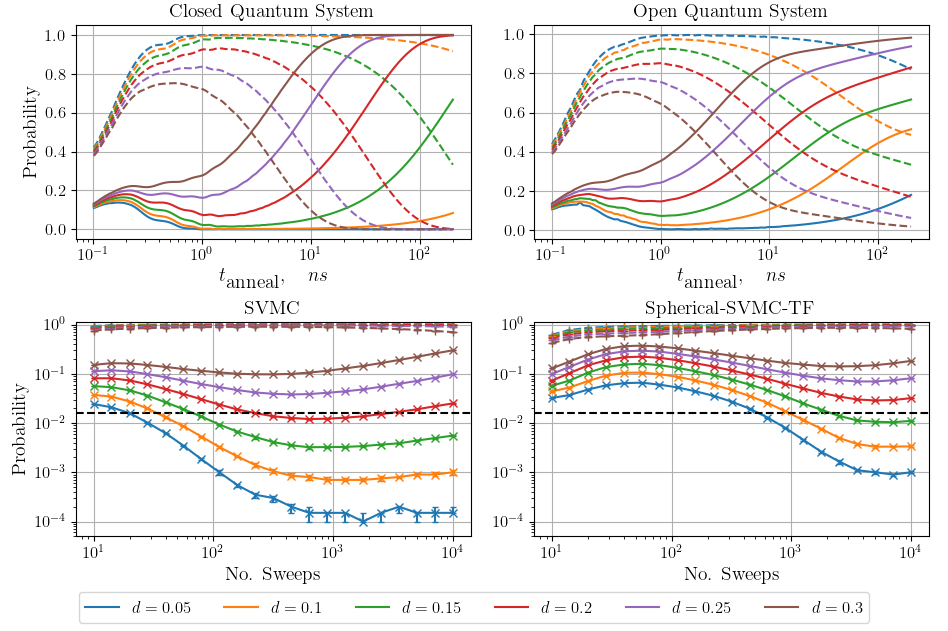}
    \caption{Plots of state probability for being in either the ground state (solid lines) or any of the $2^M$-degenerate $1^{st}$ excited states (dashed lines) at the end of an anneal. A PFC with system size $M = 3$  was evolved using quantum (top row) and classical (bottom row) dynamics. The black dashed line indicates the probability with random guessing i.e. $1/64$. The quantum simulations are plotted against anneal time in $ns$, whilst the SVMC simulations are plotted against the number of sweeps. The closed and open system dynamics are evolved according to the von-Neumann and adiabatic master equation respectively (see section~\ref{Methods_analytical}). The SVMC and spherical-SVMC-TF probabilities are found from 20,000 samples, of which we repeat 50 times and bootstrap to find a median and $ 95 \% $ confidence intervals for the data point and error bars respectively.}
    \label{fig:result_quantum_svmc_n=3}
\end{figure*}

We begin the dynamical analysis of the PFC by observing the performance of the SVMC variants when scaling in $M$. The systems chosen meet the $d < (M-1)^{-1}$ condition such that the first excited state is the exponentially large manifold which is at least a Hamming distance of $M$ away from the ground state. The combined effects of an exponentially scaling gap and manifold are observed in Fig.~\ref{fig:result_svmc_n_test} by measuring the ground state probability at the end of the anneal with respect to the number of incremental sweeps used in both SVMC and spherical-SVMC-TF. Typically we expect an increasing number of sweeps to correspond to an increasing ground state probability, but here there are three distinct regimes when annealing the PFC. For low sweep numbers, where the semi-classical potential is evolved in large steps, we have a relatively high ground state probability as the false minimum is not well resolved but SVMC still guides the spin-vector to the low-energy states. 

For medium sweep numbers we see reduction in ground state probability, caused by the quasi-continuous evolution of the semi-classical potential now leading the SVMC algorithm to the false minimum. This guides SVMC to the $\theta_b = \pi$ manifold (Fig.~\ref{fig:semi-classical}) corresponding to the degenerate computational first-excited-states, causing SVMC to spread out into this manifold and into states that are potentially further in Hamming distance from the computational ground state (see Fig.~\ref{fig:result_populations} for further evidence of this). Finally, for high sweep numbers SVMC starts to thermally equilibrate and ground state probability begins to return.

To confirm the detrimental role that the exponential manifold has on SVMC, the state probabilities just after the minimum gap are measured in Fig.~\ref{fig:result_populations} for both spherical-SVMC-TF (for $10,000$ sweeps) and a system evolved using AME (for $200 ns$). The probability distribution was measured at $s = 0.83$ for a PFC ($M = 3$, $R = 1.0$, $d = 0.1$) with a minimum gap at $s = 0.8227$. It can be seen that the probability density spreads into the computational first excited state manifold when evolving using spherical-SVMC-TF, which is caused by following the false minimum and accessing the $\theta_b = \pi$ manifold using classical dynamics. However, a system evolved using the AME remains in the lowest instantaneous eigenstates after the minimum gap, where $E_0$ also has a large overlap with the computational ground state, $\langle E_0(s=0.83)|0^{\otimes N}\rangle = 0.98$. After this point in the anneal, both the AME and spherical-SVMC-TF experience freeze-out, which prevents any more dynamical evolution that could affect ground state probability. This is illustrated by the ground state probability of the AME (spherical-SVMC-TF) evolution at $s = 0.83$ being $\sim0.50$ $(\sim3.4\times 10^{-3})$, compared to that at $s = 1$ being $\sim0.51$ $(\sim3.4\times 10^{-3})$ - see Fig.~\ref{fig:result_quantum_svmc_n=3}. It is also worth noting that we measure in the computational basis for SVMC since it is a classical algorithm with no other analogous discrete states to compare against the instantaneous states used by the AME.

Finally, we explore how the tunable hardness parameter, $d$, affects the PFC in both quantum and SVMC simulations for $M = 3$, in Fig.~\ref{fig:result_quantum_svmc_n=3}. We measure the probability at the end of the anneal of being in the ground state as well as any of the first excited states for different annealing durations. The value of $d$ also determines the size of the minimum gap, such that we span $d$ to capture various regimes at a fixed system temperature of $12$ mK. At $d \sim 0.227$ we have a minimum gap that approximately equals the system temperature.

The form of SVMC and spherical-SVMC-TF when scaling in $d$ in Fig.~\ref{fig:result_quantum_svmc_n=3} is similar to what is also seen in Fig.~\ref{fig:result_svmc_n_test} when scaling in $M$. For all sweeps, the PFC becomes harder as $d$ becomes smaller, and SVMC and spherical-SVMC-TF preferably anneal to the first excited state manifold (following the false minimum). Additionally, spherical-SVMC-TF consistently out-performs SVMC for our hardest problems ($d \leq 0.15$), something which is further discussed in appendix~\ref{Appendix_svmc_var}.

The probabilities at the end of a closed-system quantum anneal (for time $t_{\textrm{anneal}}$) are of a similar form, with a bump for short anneal times (the diabatic bump \cite{zhouQuantumApproximateOptimization2020}) and then reaching the adiabatic limit (where the ground state probability tends towards 1) at longer anneal times (e.g. $t_{\textrm{anneal}} \simeq 20 ns$ for $d = 0.3$). The closed system dynamics describe quantum evolution at a temperature of $0 K$ with no dephasing, and therefore requires a run time of $t_{\textrm{anneal}} = O(1 / \Delta_{10}^2)$ to run adiabatically. For example, at $d = 0.05$ the approximate adiabatic run time is $ \sim 163 \mu s$ and we therefore see a probability $\sim 0$ due to the short run times (see appendix~\ref{Appendix_svmc_var} for a log-scale plot of the closed-system quantum anneal).

\begin{figure}
    \centering
    \includegraphics[width=\columnwidth]{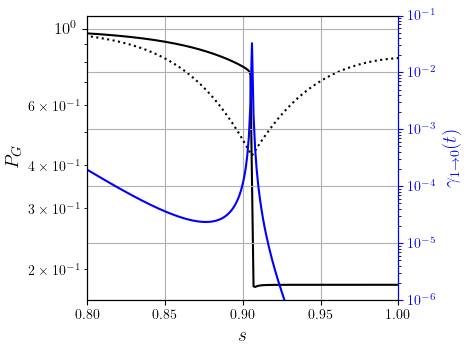}
    \caption{Plot of the evolution of ground state probability from the AME (black solid) and the Gibbs state (black dotted), as well as the transition rate (blue) (Eq.~\ref{eq:transition}). A PFC of $M = 3$, $R=1.0$, and $d = 0.05$ was simulated at a system temperature of 12 mK. The AME was evolved for $t_{\textrm{anneal}} = 200 ns$ and the minimum gap occurs at $s = 0.9059$.}
    \label{fig:result_thermal_rate}
\end{figure}

However, for the open system simulations we see a non-zero ground state probability at $d = 0.05$ for $t_{\textrm{anneal}} = 200 ns$. This can be attributed to thermally assisted quantum transitions, whereby the relaxation rate (Eq.~(\ref{eq:transition})) is non-zero about the minimum gap (Fig.~\ref{fig:result_thermal_rate}) and can return probability to the ground state from the first excited state. This can occur because the system is sub-thermal~\cite{chenWhyWhenPausing2020} (i.e. the ground-state probability is less than that at thermal equilibrium) immediately after the minimum gap, and the energy gap is still small enough to allow significant thermalization from the first excited state to the ground state. Additionally, this transition involves all backbone qubits changing their magnetization simultaneously.

Therefore thermalization in a quantum system is seen to be of some computational use, as has been seen in other literature~\cite{dicksonThermallyAssistedQuantum2013, marshallPowerPausingAdvancing2019, chenWhyWhenPausing2020, albashComparingRelaxationMechanisms2021, chancellorExperimentalTestSearch2021}. However, we see that on the time-scales tested that thermalization in classical SVMC plays a far less significant role, and results in a ground state probability (after $10,000$ sweeps) two orders of magnitude lower than a $100ns$ AME evolved anneal. The marked difference between the two types of dynamical simulations highlight the effect of the minimum gap and the exponential manifold on ground state probability, making it a gadget of interest for when differentiating between quantum and classical evolutions in the TFIM. For a better contrast between dynamics, annealing larger versions of the PFC ($M > 10$) would result in negligible statistical contribution from random state sampling, and an extremely large first excited state manifold that would likely result in SVMC failing to find the ground state. Open system simulations of this kind are unfeasible, and would most likely only be realised on experimental quantum hardware.

\section{Conclusion}

In this work, we have introduced the perturbed ferromagnetic chain (PFC), a gadget with an exponentially large first-excited-state manifold and an isolated ground state, whereby problem hardness and frustration is tuned by the perturbative parameter, $d$. When annealed in the transverse field Ising model (TFIM), the PFC develops computationally hard characteristics such as an exponentially small minimum gap (in $N$), a quantum phase transition, and a false minimum.

The evolution of the PFC in the TFIM was assessed with quantum dynamics using the adiabatic master equation (AME), and classically using both spin-vector Monte Carlo (SVMC) and spherical-SVMC-TF (see section~\ref{Methods_semiclassical} for more information). For quasi-continuous evolution of the PFC with the SVMC methods, the false minimum is followed to the computational first-excited-state manifold. This results in probable transitions to other low energy states further in Hamming distance away from the computational ground state (Fig.~\ref{fig:result_populations}), and therefore reduces the probability of reaching the ground state significantly. This is compounded by increasing problem size (Fig.~\ref{fig:result_svmc_n_test}) and by tuning $d$ (Fig.~\ref{fig:result_quantum_svmc_n=3}).

For a PFC evolved using the AME, the system mostly remains in the lowest two eigenstates (Fig.~\ref{fig:result_populations}) instead of accessing the exponential manifold, such that a $100ns$ open-quantum-system anneal results in a ground state probability two-orders of magnitude larger than a $10,000$ sweep spherical-SVMC-TF anneal for the hardest comparative problem simulated (Fig.~\ref{fig:result_quantum_svmc_n=3}). The AME evolution permits thermalization to the ground state (Fig.~\ref{fig:result_thermal_rate}) in time-scales too short for adiabatic evolution, indicating that thermalization aids computation of the PFC in the TFIM.

The distinct differences seen between classical and quantum evolutions therefore makes the PFC a useful gadget in differentiating dynamical systems, something which other gadgets cannot always exhibit~\cite{albashDemonstrationScalingAdvantage2018, muthukrishnanTunnelingSpeedupQuantum2016, albashComparingRelaxationMechanisms2021}. Comparative tests at larger PFC system sizes of $M > 10$ would also allow for better differentiation and insight into the computational extent of thermalization in quantum and classical evolutions. However, realizing open quantum system simulations of this scale is intractable, and would therefore require further approximations or an experimental implementation with quantum hardware.  \\

\section*{Acknowledgements}

We gratefully acknowledge Catherine White, Gioele Consani, Robert Banks and Nedeen Alsharif for useful comments and insightful discussion. The research is based upon work (partially) supported by the Office of the Director of National Intelligence (ODNI), Intelligence Advanced Research Projects Activity (IARPA), and the Defense Advanced Research Projects Agency (DARPA), via the U.S. Army Research Office Contract No. W911NF-17-C-0050. The views and conclusions contained herein are those of the authors and should not be interpreted as necessarily representing the official policies or endorsements, either expressed or implied, of the ODNI, IARPA, DARPA, or the U.S. Government. The U.S. Government is authorized to reproduce and distribute reprints for Governmental purposes notwithstanding any copyright annotation thereon. D. T. O’C. thanks EPSRC and BT for additional support. 

\medskip
\bibliography{main}

\begin{thebibliography}{39}%
\makeatletter
\providecommand \@ifxundefined [1]{%
 \@ifx{#1\undefined}
}%
\providecommand \@ifnum [1]{%
 \ifnum #1\expandafter \@firstoftwo
 \else \expandafter \@secondoftwo
 \fi
}%
\providecommand \@ifx [1]{%
 \ifx #1\expandafter \@firstoftwo
 \else \expandafter \@secondoftwo
 \fi
}%
\providecommand \natexlab [1]{#1}%
\providecommand \enquote  [1]{``#1''}%
\providecommand \bibnamefont  [1]{#1}%
\providecommand \bibfnamefont [1]{#1}%
\providecommand \citenamefont [1]{#1}%
\providecommand \href@noop [0]{\@secondoftwo}%
\providecommand \href [0]{\begingroup \@sanitize@url \@href}%
\providecommand \@href[1]{\@@startlink{#1}\@@href}%
\providecommand \@@href[1]{\endgroup#1\@@endlink}%
\providecommand \@sanitize@url [0]{\catcode `\\12\catcode `\$12\catcode
  `\&12\catcode `\#12\catcode `\^12\catcode `\_12\catcode `\%12\relax}%
\providecommand \@@startlink[1]{}%
\providecommand \@@endlink[0]{}%
\providecommand \url  [0]{\begingroup\@sanitize@url \@url }%
\providecommand \@url [1]{\endgroup\@href {#1}{\urlprefix }}%
\providecommand \urlprefix  [0]{URL }%
\providecommand \Eprint [0]{\href }%
\providecommand \doibase [0]{https://doi.org/}%
\providecommand \selectlanguage [0]{\@gobble}%
\providecommand \bibinfo  [0]{\@secondoftwo}%
\providecommand \bibfield  [0]{\@secondoftwo}%
\providecommand \translation [1]{[#1]}%
\providecommand \BibitemOpen [0]{}%
\providecommand \bibitemStop [0]{}%
\providecommand \bibitemNoStop [0]{.\EOS\space}%
\providecommand \EOS [0]{\spacefactor3000\relax}%
\providecommand \BibitemShut  [1]{\csname bibitem#1\endcsname}%
\let\auto@bib@innerbib\@empty
\bibitem [{\citenamefont {Nagaj}\ \emph {et~al.}(2012)\citenamefont {Nagaj},
  \citenamefont {Somma},\ and\ \citenamefont
  {Kieferova}}]{nagajQuantumSpeedupQuantum2012}%
  \BibitemOpen
  \bibfield  {author} {\bibinfo {author} {\bibfnamefont {D.}~\bibnamefont
  {Nagaj}}, \bibinfo {author} {\bibfnamefont {R.~D.}\ \bibnamefont {Somma}},\
  and\ \bibinfo {author} {\bibfnamefont {M.}~\bibnamefont {Kieferova}},\
  }\bibfield  {title} {\bibinfo {title} {Quantum {{Speedup}} by {{Quantum
  Annealing}}},\ }\href {https://doi.org/10.1103/PhysRevLett.109.050501}
  {\bibfield  {journal} {\bibinfo  {journal} {Physical Review Letters}\
  }\textbf {\bibinfo {volume} {109}},\ \bibinfo {pages} {050501} (\bibinfo
  {year} {2012})},\ \Eprint {https://arxiv.org/abs/1202.6257} {arXiv:1202.6257}
  \BibitemShut {NoStop}%
\bibitem [{\citenamefont {Denchev}\ \emph {et~al.}(2016)\citenamefont
  {Denchev}, \citenamefont {Boixo}, \citenamefont {Isakov}, \citenamefont
  {Ding}, \citenamefont {Babbush}, \citenamefont {Smelyanskiy}, \citenamefont
  {Martinis},\ and\ \citenamefont {Neven}}]{denchevWhatComputationalValue2016}%
  \BibitemOpen
  \bibfield  {author} {\bibinfo {author} {\bibfnamefont {V.~S.}\ \bibnamefont
  {Denchev}}, \bibinfo {author} {\bibfnamefont {S.}~\bibnamefont {Boixo}},
  \bibinfo {author} {\bibfnamefont {S.~V.}\ \bibnamefont {Isakov}}, \bibinfo
  {author} {\bibfnamefont {N.}~\bibnamefont {Ding}}, \bibinfo {author}
  {\bibfnamefont {R.}~\bibnamefont {Babbush}}, \bibinfo {author} {\bibfnamefont
  {V.}~\bibnamefont {Smelyanskiy}}, \bibinfo {author} {\bibfnamefont
  {J.}~\bibnamefont {Martinis}},\ and\ \bibinfo {author} {\bibfnamefont
  {H.}~\bibnamefont {Neven}},\ }\bibfield  {title} {\bibinfo {title} {What is
  the {{Computational Value}} of {{Finite}}-{{Range Tunneling}}?},\ }\href
  {https://doi.org/10.1103/PhysRevX.6.031015} {\bibfield  {journal} {\bibinfo
  {journal} {Physical Review X}\ }\textbf {\bibinfo {volume} {6}},\ \bibinfo
  {pages} {031015} (\bibinfo {year} {2016})}\BibitemShut {NoStop}%
\bibitem [{\citenamefont {Boixo}\ \emph {et~al.}(2016)\citenamefont {Boixo},
  \citenamefont {Smelyanskiy}, \citenamefont {Shabani}, \citenamefont {Isakov},
  \citenamefont {Dykman}, \citenamefont {Denchev}, \citenamefont {Amin},
  \citenamefont {Smirnov}, \citenamefont {Mohseni},\ and\ \citenamefont
  {Neven}}]{boixoComputationalMultiqubitTunnelling2016}%
  \BibitemOpen
  \bibfield  {author} {\bibinfo {author} {\bibfnamefont {S.}~\bibnamefont
  {Boixo}}, \bibinfo {author} {\bibfnamefont {V.~N.}\ \bibnamefont
  {Smelyanskiy}}, \bibinfo {author} {\bibfnamefont {A.}~\bibnamefont
  {Shabani}}, \bibinfo {author} {\bibfnamefont {S.~V.}\ \bibnamefont {Isakov}},
  \bibinfo {author} {\bibfnamefont {M.}~\bibnamefont {Dykman}}, \bibinfo
  {author} {\bibfnamefont {V.~S.}\ \bibnamefont {Denchev}}, \bibinfo {author}
  {\bibfnamefont {M.~H.}\ \bibnamefont {Amin}}, \bibinfo {author}
  {\bibfnamefont {A.~Y.}\ \bibnamefont {Smirnov}}, \bibinfo {author}
  {\bibfnamefont {M.}~\bibnamefont {Mohseni}},\ and\ \bibinfo {author}
  {\bibfnamefont {H.}~\bibnamefont {Neven}},\ }\bibfield  {title} {\bibinfo
  {title} {Computational multiqubit tunnelling in programmable quantum
  annealers},\ }\href {https://doi.org/10.1038/ncomms10327} {\bibfield
  {journal} {\bibinfo  {journal} {Nature Communications}\ }\textbf {\bibinfo
  {volume} {7}},\ \bibinfo {pages} {10327} (\bibinfo {year}
  {2016})}\BibitemShut {NoStop}%
\bibitem [{\citenamefont {Albash}\ and\ \citenamefont
  {Lidar}(2018)}]{albashDemonstrationScalingAdvantage2018}%
  \BibitemOpen
  \bibfield  {author} {\bibinfo {author} {\bibfnamefont {T.}~\bibnamefont
  {Albash}}\ and\ \bibinfo {author} {\bibfnamefont {D.~A.}\ \bibnamefont
  {Lidar}},\ }\bibfield  {title} {\bibinfo {title} {Demonstration of a scaling
  advantage for a quantum annealer over simulated annealing},\ }\href
  {https://doi.org/10.1103/PhysRevX.8.031016} {\bibfield  {journal} {\bibinfo
  {journal} {Physical Review X}\ }\textbf {\bibinfo {volume} {8}},\ \bibinfo
  {pages} {031016} (\bibinfo {year} {2018})},\ \Eprint
  {https://arxiv.org/abs/1705.07452} {arXiv:1705.07452} \BibitemShut {NoStop}%
\bibitem [{\citenamefont {Mandr{\`a}}\ and\ \citenamefont
  {Katzgraber}(2018)}]{mandraDeceptiveStepQuantum2018}%
  \BibitemOpen
  \bibfield  {author} {\bibinfo {author} {\bibfnamefont {S.}~\bibnamefont
  {Mandr{\`a}}}\ and\ \bibinfo {author} {\bibfnamefont {H.~G.}\ \bibnamefont
  {Katzgraber}},\ }\bibfield  {title} {\bibinfo {title} {A deceptive step
  towards quantum speedup detection},\ }\href
  {https://doi.org/10.1088/2058-9565/aac8b2} {\bibfield  {journal} {\bibinfo
  {journal} {Quantum Science and Technology}\ }\textbf {\bibinfo {volume}
  {3}},\ \bibinfo {pages} {04LT01} (\bibinfo {year} {2018})}\BibitemShut
  {NoStop}%
\bibitem [{\citenamefont {King}\ \emph {et~al.}(2021)\citenamefont {King},
  \citenamefont {Raymond}, \citenamefont {Lanting}, \citenamefont {Isakov},
  \citenamefont {Mohseni}, \citenamefont {{Poulin-Lamarre}}, \citenamefont
  {Ejtemaee}, \citenamefont {Bernoudy}, \citenamefont {Ozfidan}, \citenamefont
  {Smirnov}, \citenamefont {Reis}, \citenamefont {Altomare}, \citenamefont
  {Babcock}, \citenamefont {Baron}, \citenamefont {Berkley}, \citenamefont
  {Boothby}, \citenamefont {Bunyk}, \citenamefont {Christiani}, \citenamefont
  {Enderud}, \citenamefont {Evert}, \citenamefont {Harris}, \citenamefont
  {Hoskinson}, \citenamefont {Huang}, \citenamefont {Jooya}, \citenamefont
  {Khodabandelou}, \citenamefont {Ladizinsky}, \citenamefont {Li},
  \citenamefont {Lott}, \citenamefont {MacDonald}, \citenamefont {Marsden},
  \citenamefont {Marsden}, \citenamefont {Medina}, \citenamefont {Molavi},
  \citenamefont {Neufeld}, \citenamefont {Norouzpour}, \citenamefont {Oh},
  \citenamefont {Pavlov}, \citenamefont {Perminov}, \citenamefont {Prescott},
  \citenamefont {Rich}, \citenamefont {Sato}, \citenamefont {Sheldan},
  \citenamefont {Sterling}, \citenamefont {Swenson}, \citenamefont {Tsai},
  \citenamefont {Volkmann}, \citenamefont {Whittaker}, \citenamefont
  {Wilkinson}, \citenamefont {Yao}, \citenamefont {Neven}, \citenamefont
  {Hilton}, \citenamefont {Ladizinsky}, \citenamefont {Johnson},\ and\
  \citenamefont {Amin}}]{kingScalingAdvantagePathintegral2021}%
  \BibitemOpen
  \bibfield  {author} {\bibinfo {author} {\bibfnamefont {A.~D.}\ \bibnamefont
  {King}}, \bibinfo {author} {\bibfnamefont {J.}~\bibnamefont {Raymond}},
  \bibinfo {author} {\bibfnamefont {T.}~\bibnamefont {Lanting}}, \bibinfo
  {author} {\bibfnamefont {S.~V.}\ \bibnamefont {Isakov}}, \bibinfo {author}
  {\bibfnamefont {M.}~\bibnamefont {Mohseni}}, \bibinfo {author} {\bibfnamefont
  {G.}~\bibnamefont {{Poulin-Lamarre}}}, \bibinfo {author} {\bibfnamefont
  {S.}~\bibnamefont {Ejtemaee}}, \bibinfo {author} {\bibfnamefont
  {W.}~\bibnamefont {Bernoudy}}, \bibinfo {author} {\bibfnamefont
  {I.}~\bibnamefont {Ozfidan}}, \bibinfo {author} {\bibfnamefont {A.~Y.}\
  \bibnamefont {Smirnov}}, \bibinfo {author} {\bibfnamefont {M.}~\bibnamefont
  {Reis}}, \bibinfo {author} {\bibfnamefont {F.}~\bibnamefont {Altomare}},
  \bibinfo {author} {\bibfnamefont {M.}~\bibnamefont {Babcock}}, \bibinfo
  {author} {\bibfnamefont {C.}~\bibnamefont {Baron}}, \bibinfo {author}
  {\bibfnamefont {A.~J.}\ \bibnamefont {Berkley}}, \bibinfo {author}
  {\bibfnamefont {K.}~\bibnamefont {Boothby}}, \bibinfo {author} {\bibfnamefont
  {P.~I.}\ \bibnamefont {Bunyk}}, \bibinfo {author} {\bibfnamefont
  {H.}~\bibnamefont {Christiani}}, \bibinfo {author} {\bibfnamefont
  {C.}~\bibnamefont {Enderud}}, \bibinfo {author} {\bibfnamefont
  {B.}~\bibnamefont {Evert}}, \bibinfo {author} {\bibfnamefont
  {R.}~\bibnamefont {Harris}}, \bibinfo {author} {\bibfnamefont
  {E.}~\bibnamefont {Hoskinson}}, \bibinfo {author} {\bibfnamefont
  {S.}~\bibnamefont {Huang}}, \bibinfo {author} {\bibfnamefont
  {K.}~\bibnamefont {Jooya}}, \bibinfo {author} {\bibfnamefont
  {A.}~\bibnamefont {Khodabandelou}}, \bibinfo {author} {\bibfnamefont
  {N.}~\bibnamefont {Ladizinsky}}, \bibinfo {author} {\bibfnamefont
  {R.}~\bibnamefont {Li}}, \bibinfo {author} {\bibfnamefont {P.~A.}\
  \bibnamefont {Lott}}, \bibinfo {author} {\bibfnamefont {A.~J.~R.}\
  \bibnamefont {MacDonald}}, \bibinfo {author} {\bibfnamefont {D.}~\bibnamefont
  {Marsden}}, \bibinfo {author} {\bibfnamefont {G.}~\bibnamefont {Marsden}},
  \bibinfo {author} {\bibfnamefont {T.}~\bibnamefont {Medina}}, \bibinfo
  {author} {\bibfnamefont {R.}~\bibnamefont {Molavi}}, \bibinfo {author}
  {\bibfnamefont {R.}~\bibnamefont {Neufeld}}, \bibinfo {author} {\bibfnamefont
  {M.}~\bibnamefont {Norouzpour}}, \bibinfo {author} {\bibfnamefont
  {T.}~\bibnamefont {Oh}}, \bibinfo {author} {\bibfnamefont {I.}~\bibnamefont
  {Pavlov}}, \bibinfo {author} {\bibfnamefont {I.}~\bibnamefont {Perminov}},
  \bibinfo {author} {\bibfnamefont {T.}~\bibnamefont {Prescott}}, \bibinfo
  {author} {\bibfnamefont {C.}~\bibnamefont {Rich}}, \bibinfo {author}
  {\bibfnamefont {Y.}~\bibnamefont {Sato}}, \bibinfo {author} {\bibfnamefont
  {B.}~\bibnamefont {Sheldan}}, \bibinfo {author} {\bibfnamefont
  {G.}~\bibnamefont {Sterling}}, \bibinfo {author} {\bibfnamefont {L.~J.}\
  \bibnamefont {Swenson}}, \bibinfo {author} {\bibfnamefont {N.}~\bibnamefont
  {Tsai}}, \bibinfo {author} {\bibfnamefont {M.~H.}\ \bibnamefont {Volkmann}},
  \bibinfo {author} {\bibfnamefont {J.~D.}\ \bibnamefont {Whittaker}}, \bibinfo
  {author} {\bibfnamefont {W.}~\bibnamefont {Wilkinson}}, \bibinfo {author}
  {\bibfnamefont {J.}~\bibnamefont {Yao}}, \bibinfo {author} {\bibfnamefont
  {H.}~\bibnamefont {Neven}}, \bibinfo {author} {\bibfnamefont {J.~P.}\
  \bibnamefont {Hilton}}, \bibinfo {author} {\bibfnamefont {E.}~\bibnamefont
  {Ladizinsky}}, \bibinfo {author} {\bibfnamefont {M.~W.}\ \bibnamefont
  {Johnson}},\ and\ \bibinfo {author} {\bibfnamefont {M.~H.}\ \bibnamefont
  {Amin}},\ }\bibfield  {title} {\bibinfo {title} {Scaling advantage over
  path-integral {{Monte Carlo}} in quantum simulation of geometrically
  frustrated magnets},\ }\href {https://doi.org/10.1038/s41467-021-20901-5}
  {\bibfield  {journal} {\bibinfo  {journal} {Nature Communications}\ }\textbf
  {\bibinfo {volume} {12}},\ \bibinfo {pages} {1113} (\bibinfo {year}
  {2021})}\BibitemShut {NoStop}%
\bibitem [{\citenamefont {Kadowaki}\ and\ \citenamefont
  {Nishimori}(1998)}]{kadowakiQuantumAnnealingTransverse1998}%
  \BibitemOpen
  \bibfield  {author} {\bibinfo {author} {\bibfnamefont {T.}~\bibnamefont
  {Kadowaki}}\ and\ \bibinfo {author} {\bibfnamefont {H.}~\bibnamefont
  {Nishimori}},\ }\bibfield  {title} {\bibinfo {title} {Quantum annealing in
  the transverse {{Ising}} model},\ }\href
  {https://doi.org/10.1103/PhysRevE.58.5355} {\bibfield  {journal} {\bibinfo
  {journal} {Physical Review E}\ }\textbf {\bibinfo {volume} {58}},\ \bibinfo
  {pages} {5355} (\bibinfo {year} {1998})}\BibitemShut {NoStop}%
\bibitem [{\citenamefont {Farhi}\ \emph {et~al.}(2001)\citenamefont {Farhi},
  \citenamefont {Goldstone}, \citenamefont {Gutmann}, \citenamefont {Lapan},
  \citenamefont {Lundgren},\ and\ \citenamefont
  {Preda}}]{farhiQuantumAdiabaticEvolution2001}%
  \BibitemOpen
  \bibfield  {author} {\bibinfo {author} {\bibfnamefont {E.}~\bibnamefont
  {Farhi}}, \bibinfo {author} {\bibfnamefont {J.}~\bibnamefont {Goldstone}},
  \bibinfo {author} {\bibfnamefont {S.}~\bibnamefont {Gutmann}}, \bibinfo
  {author} {\bibfnamefont {J.}~\bibnamefont {Lapan}}, \bibinfo {author}
  {\bibfnamefont {A.}~\bibnamefont {Lundgren}},\ and\ \bibinfo {author}
  {\bibfnamefont {D.}~\bibnamefont {Preda}},\ }\bibfield  {title} {\bibinfo
  {title} {A {{Quantum Adiabatic Evolution Algorithm Applied}} to {{Random
  Instances}} of an {{NP}}-{{Complete Problem}}},\ }\href
  {https://doi.org/10.1126/science.1057726} {\bibfield  {journal} {\bibinfo
  {journal} {Science}\ }\textbf {\bibinfo {volume} {292}},\ \bibinfo {pages}
  {472} (\bibinfo {year} {2001})}\BibitemShut {NoStop}%
\bibitem [{\citenamefont {Venturelli}\ and\ \citenamefont
  {Kondratyev}(2019)}]{venturelliReverseQuantumAnnealing2019}%
  \BibitemOpen
  \bibfield  {author} {\bibinfo {author} {\bibfnamefont {D.}~\bibnamefont
  {Venturelli}}\ and\ \bibinfo {author} {\bibfnamefont {A.}~\bibnamefont
  {Kondratyev}},\ }\bibfield  {title} {\bibinfo {title} {Reverse {{Quantum
  Annealing Approach}} to {{Portfolio Optimization Problems}}},\ }\href
  {https://doi.org/10.1007/s42484-019-00001-w} {\bibfield  {journal} {\bibinfo
  {journal} {Quantum Machine Intelligence}\ }\textbf {\bibinfo {volume} {1}},\
  \bibinfo {pages} {17} (\bibinfo {year} {2019})},\ \Eprint
  {https://arxiv.org/abs/1810.08584} {arXiv:1810.08584} \BibitemShut {NoStop}%
\bibitem [{\citenamefont {Stollenwerk}\ \emph {et~al.}(2018)\citenamefont
  {Stollenwerk}, \citenamefont {Lobe},\ and\ \citenamefont
  {Jung}}]{stollenwerkFlightGateAssignment2018}%
  \BibitemOpen
  \bibfield  {author} {\bibinfo {author} {\bibfnamefont {T.}~\bibnamefont
  {Stollenwerk}}, \bibinfo {author} {\bibfnamefont {E.}~\bibnamefont {Lobe}},\
  and\ \bibinfo {author} {\bibfnamefont {M.}~\bibnamefont {Jung}},\ }\bibfield
  {title} {\bibinfo {title} {Flight {{Gate Assignment}} with a {{Quantum
  Annealer}}},\ }\href@noop {} {\bibfield  {journal} {\bibinfo  {journal}
  {arXiv:1811.09465 [quant-ph]}\ } (\bibinfo {year} {2018})},\ \Eprint
  {https://arxiv.org/abs/1811.09465} {arXiv:1811.09465 [quant-ph]} \BibitemShut
  {NoStop}%
\bibitem [{\citenamefont {Kim}\ \emph {et~al.}(2019)\citenamefont {Kim},
  \citenamefont {Venturelli},\ and\ \citenamefont
  {Jamieson}}]{kimLeveragingQuantumAnnealing2019}%
  \BibitemOpen
  \bibfield  {author} {\bibinfo {author} {\bibfnamefont {M.}~\bibnamefont
  {Kim}}, \bibinfo {author} {\bibfnamefont {D.}~\bibnamefont {Venturelli}},\
  and\ \bibinfo {author} {\bibfnamefont {K.}~\bibnamefont {Jamieson}},\
  }\bibfield  {title} {\bibinfo {title} {Leveraging {{Quantum Annealing}} for
  {{Large MIMO Processing}} in {{Centralized Radio Access Networks}}},\ }\href
  {https://doi.org/10.1145/3341302.3342072} {\bibfield  {journal} {\bibinfo
  {journal} {Proceedings of the ACM Special Interest Group on Data
  Communication}\ ,\ \bibinfo {pages} {241}} (\bibinfo {year} {2019})},\
  \Eprint {https://arxiv.org/abs/2001.04014} {arXiv:2001.04014} \BibitemShut
  {NoStop}%
\bibitem [{\citenamefont {Inoue}\ \emph {et~al.}(2021)\citenamefont {Inoue},
  \citenamefont {Okada}, \citenamefont {Matsumori}, \citenamefont {Aihara},\
  and\ \citenamefont {Yoshida}}]{inoueTrafficSignalOptimization2021}%
  \BibitemOpen
  \bibfield  {author} {\bibinfo {author} {\bibfnamefont {D.}~\bibnamefont
  {Inoue}}, \bibinfo {author} {\bibfnamefont {A.}~\bibnamefont {Okada}},
  \bibinfo {author} {\bibfnamefont {T.}~\bibnamefont {Matsumori}}, \bibinfo
  {author} {\bibfnamefont {K.}~\bibnamefont {Aihara}},\ and\ \bibinfo {author}
  {\bibfnamefont {H.}~\bibnamefont {Yoshida}},\ }\bibfield  {title} {\bibinfo
  {title} {Traffic signal optimization on a square lattice with quantum
  annealing},\ }\href {https://doi.org/10.1038/s41598-021-82740-0} {\bibfield
  {journal} {\bibinfo  {journal} {Scientific Reports}\ }\textbf {\bibinfo
  {volume} {11}},\ \bibinfo {pages} {3303} (\bibinfo {year}
  {2021})}\BibitemShut {NoStop}%
\bibitem [{\citenamefont {Kitai}\ \emph {et~al.}(2020)\citenamefont {Kitai},
  \citenamefont {Guo}, \citenamefont {Ju}, \citenamefont {Tanaka},
  \citenamefont {Tsuda}, \citenamefont {Shiomi},\ and\ \citenamefont
  {Tamura}}]{kitaiDesigningMetamaterialsQuantum2020}%
  \BibitemOpen
  \bibfield  {author} {\bibinfo {author} {\bibfnamefont {K.}~\bibnamefont
  {Kitai}}, \bibinfo {author} {\bibfnamefont {J.}~\bibnamefont {Guo}}, \bibinfo
  {author} {\bibfnamefont {S.}~\bibnamefont {Ju}}, \bibinfo {author}
  {\bibfnamefont {S.}~\bibnamefont {Tanaka}}, \bibinfo {author} {\bibfnamefont
  {K.}~\bibnamefont {Tsuda}}, \bibinfo {author} {\bibfnamefont
  {J.}~\bibnamefont {Shiomi}},\ and\ \bibinfo {author} {\bibfnamefont
  {R.}~\bibnamefont {Tamura}},\ }\bibfield  {title} {\bibinfo {title}
  {Designing metamaterials with quantum annealing and factorization machines},\
  }\href {https://doi.org/10.1103/PhysRevResearch.2.013319} {\bibfield
  {journal} {\bibinfo  {journal} {Physical Review Research}\ }\textbf {\bibinfo
  {volume} {2}},\ \bibinfo {pages} {013319} (\bibinfo {year}
  {2020})}\BibitemShut {NoStop}%
\bibitem [{\citenamefont {Asproni}\ \emph {et~al.}(2020)\citenamefont
  {Asproni}, \citenamefont {Caputo}, \citenamefont {Silva}, \citenamefont
  {Fazzi},\ and\ \citenamefont
  {Magagnini}}]{asproniAccuracyMinorEmbedding2020}%
  \BibitemOpen
  \bibfield  {author} {\bibinfo {author} {\bibfnamefont {L.}~\bibnamefont
  {Asproni}}, \bibinfo {author} {\bibfnamefont {D.}~\bibnamefont {Caputo}},
  \bibinfo {author} {\bibfnamefont {B.}~\bibnamefont {Silva}}, \bibinfo
  {author} {\bibfnamefont {G.}~\bibnamefont {Fazzi}},\ and\ \bibinfo {author}
  {\bibfnamefont {M.}~\bibnamefont {Magagnini}},\ }\bibfield  {title} {\bibinfo
  {title} {Accuracy and minor embedding in subqubo decomposition with fully
  connected large problems: A case study about the number partitioning
  problem},\ }\href {https://doi.org/10.1007/s42484-020-00014-w} {\bibfield
  {journal} {\bibinfo  {journal} {Quantum Machine Intelligence}\ }\textbf
  {\bibinfo {volume} {2}},\ \bibinfo {pages} {1} (\bibinfo {year}
  {2020})}\BibitemShut {NoStop}%
\bibitem [{\citenamefont {Neukart}\ \emph {et~al.}(2017)\citenamefont
  {Neukart}, \citenamefont {Compostella}, \citenamefont {Seidel}, \citenamefont
  {{von Dollen}}, \citenamefont {Yarkoni},\ and\ \citenamefont
  {Parney}}]{neukartTrafficFlowOptimization2017}%
  \BibitemOpen
  \bibfield  {author} {\bibinfo {author} {\bibfnamefont {F.}~\bibnamefont
  {Neukart}}, \bibinfo {author} {\bibfnamefont {G.}~\bibnamefont
  {Compostella}}, \bibinfo {author} {\bibfnamefont {C.}~\bibnamefont {Seidel}},
  \bibinfo {author} {\bibfnamefont {D.}~\bibnamefont {{von Dollen}}}, \bibinfo
  {author} {\bibfnamefont {S.}~\bibnamefont {Yarkoni}},\ and\ \bibinfo {author}
  {\bibfnamefont {B.}~\bibnamefont {Parney}},\ }\bibfield  {title} {\bibinfo
  {title} {Traffic {{Flow Optimization Using}} a {{Quantum Annealer}}},\
  }\bibfield  {journal} {\bibinfo  {journal} {Frontiers in ICT}\ }\textbf
  {\bibinfo {volume} {4}},\ \href {https://doi.org/10.3389/fict.2017.00029}
  {10.3389/fict.2017.00029} (\bibinfo {year} {2017})\BibitemShut {NoStop}%
\bibitem [{\citenamefont {King}\ \emph {et~al.}(2018)\citenamefont {King},
  \citenamefont {Carrasquilla}, \citenamefont {Raymond}, \citenamefont
  {Ozfidan}, \citenamefont {Andriyash}, \citenamefont {Berkley}, \citenamefont
  {Reis}, \citenamefont {Lanting}, \citenamefont {Harris}, \citenamefont
  {Altomare}, \citenamefont {Boothby}, \citenamefont {Bunyk}, \citenamefont
  {Enderud}, \citenamefont {Fr{\'e}chette}, \citenamefont {Hoskinson},
  \citenamefont {Ladizinsky}, \citenamefont {Oh}, \citenamefont
  {{Poulin-Lamarre}}, \citenamefont {Rich}, \citenamefont {Sato}, \citenamefont
  {Smirnov}, \citenamefont {Swenson}, \citenamefont {Volkmann}, \citenamefont
  {Whittaker}, \citenamefont {Yao}, \citenamefont {Ladizinsky}, \citenamefont
  {Johnson}, \citenamefont {Hilton},\ and\ \citenamefont
  {Amin}}]{kingObservationTopologicalPhenomena2018}%
  \BibitemOpen
  \bibfield  {author} {\bibinfo {author} {\bibfnamefont {A.~D.}\ \bibnamefont
  {King}}, \bibinfo {author} {\bibfnamefont {J.}~\bibnamefont {Carrasquilla}},
  \bibinfo {author} {\bibfnamefont {J.}~\bibnamefont {Raymond}}, \bibinfo
  {author} {\bibfnamefont {I.}~\bibnamefont {Ozfidan}}, \bibinfo {author}
  {\bibfnamefont {E.}~\bibnamefont {Andriyash}}, \bibinfo {author}
  {\bibfnamefont {A.}~\bibnamefont {Berkley}}, \bibinfo {author} {\bibfnamefont
  {M.}~\bibnamefont {Reis}}, \bibinfo {author} {\bibfnamefont {T.}~\bibnamefont
  {Lanting}}, \bibinfo {author} {\bibfnamefont {R.}~\bibnamefont {Harris}},
  \bibinfo {author} {\bibfnamefont {F.}~\bibnamefont {Altomare}}, \bibinfo
  {author} {\bibfnamefont {K.}~\bibnamefont {Boothby}}, \bibinfo {author}
  {\bibfnamefont {P.~I.}\ \bibnamefont {Bunyk}}, \bibinfo {author}
  {\bibfnamefont {C.}~\bibnamefont {Enderud}}, \bibinfo {author} {\bibfnamefont
  {A.}~\bibnamefont {Fr{\'e}chette}}, \bibinfo {author} {\bibfnamefont
  {E.}~\bibnamefont {Hoskinson}}, \bibinfo {author} {\bibfnamefont
  {N.}~\bibnamefont {Ladizinsky}}, \bibinfo {author} {\bibfnamefont
  {T.}~\bibnamefont {Oh}}, \bibinfo {author} {\bibfnamefont {G.}~\bibnamefont
  {{Poulin-Lamarre}}}, \bibinfo {author} {\bibfnamefont {C.}~\bibnamefont
  {Rich}}, \bibinfo {author} {\bibfnamefont {Y.}~\bibnamefont {Sato}}, \bibinfo
  {author} {\bibfnamefont {A.~Y.}\ \bibnamefont {Smirnov}}, \bibinfo {author}
  {\bibfnamefont {L.~J.}\ \bibnamefont {Swenson}}, \bibinfo {author}
  {\bibfnamefont {M.~H.}\ \bibnamefont {Volkmann}}, \bibinfo {author}
  {\bibfnamefont {J.}~\bibnamefont {Whittaker}}, \bibinfo {author}
  {\bibfnamefont {J.}~\bibnamefont {Yao}}, \bibinfo {author} {\bibfnamefont
  {E.}~\bibnamefont {Ladizinsky}}, \bibinfo {author} {\bibfnamefont {M.~W.}\
  \bibnamefont {Johnson}}, \bibinfo {author} {\bibfnamefont {J.}~\bibnamefont
  {Hilton}},\ and\ \bibinfo {author} {\bibfnamefont {M.~H.}\ \bibnamefont
  {Amin}},\ }\bibfield  {title} {\bibinfo {title} {Observation of topological
  phenomena in a programmable lattice of 1,800 qubits},\ }\href
  {https://doi.org/10.1038/s41586-018-0410-x} {\bibfield  {journal} {\bibinfo
  {journal} {Nature}\ }\textbf {\bibinfo {volume} {560}},\ \bibinfo {pages}
  {456} (\bibinfo {year} {2018})}\BibitemShut {NoStop}%
\bibitem [{\citenamefont {Bando}\ \emph {et~al.}(2020)\citenamefont {Bando},
  \citenamefont {Susa}, \citenamefont {Oshiyama}, \citenamefont {Shibata},
  \citenamefont {Ohzeki}, \citenamefont {{G{\'o}mez-Ruiz}}, \citenamefont
  {Lidar}, \citenamefont {Suzuki}, \citenamefont {{del Campo}},\ and\
  \citenamefont {Nishimori}}]{bandoProbingUniversalityTopological2020}%
  \BibitemOpen
  \bibfield  {author} {\bibinfo {author} {\bibfnamefont {Y.}~\bibnamefont
  {Bando}}, \bibinfo {author} {\bibfnamefont {Y.}~\bibnamefont {Susa}},
  \bibinfo {author} {\bibfnamefont {H.}~\bibnamefont {Oshiyama}}, \bibinfo
  {author} {\bibfnamefont {N.}~\bibnamefont {Shibata}}, \bibinfo {author}
  {\bibfnamefont {M.}~\bibnamefont {Ohzeki}}, \bibinfo {author} {\bibfnamefont
  {F.~J.}\ \bibnamefont {{G{\'o}mez-Ruiz}}}, \bibinfo {author} {\bibfnamefont
  {D.~A.}\ \bibnamefont {Lidar}}, \bibinfo {author} {\bibfnamefont
  {S.}~\bibnamefont {Suzuki}}, \bibinfo {author} {\bibfnamefont
  {A.}~\bibnamefont {{del Campo}}},\ and\ \bibinfo {author} {\bibfnamefont
  {H.}~\bibnamefont {Nishimori}},\ }\bibfield  {title} {\bibinfo {title}
  {Probing the universality of topological defect formation in a quantum
  annealer: Kibble-{{Zurek}} mechanism and beyond},\ }\href
  {https://doi.org/10.1103/PhysRevResearch.2.033369} {\bibfield  {journal}
  {\bibinfo  {journal} {Physical Review Research}\ }\textbf {\bibinfo {volume}
  {2}},\ \bibinfo {pages} {033369} (\bibinfo {year} {2020})}\BibitemShut
  {NoStop}%
\bibitem [{\citenamefont {Harris}\ \emph {et~al.}(2008)\citenamefont {Harris},
  \citenamefont {Johnson}, \citenamefont {Han}, \citenamefont {Berkley},
  \citenamefont {Johansson}, \citenamefont {Bunyk}, \citenamefont {Ladizinsky},
  \citenamefont {Govorkov}, \citenamefont {Thom}, \citenamefont {Uchaikin},
  \citenamefont {Bumble}, \citenamefont {Fung}, \citenamefont {Kaul},
  \citenamefont {Kleinsasser}, \citenamefont {Amin},\ and\ \citenamefont
  {Averin}}]{harrisProbingNoiseFlux2008}%
  \BibitemOpen
  \bibfield  {author} {\bibinfo {author} {\bibfnamefont {R.}~\bibnamefont
  {Harris}}, \bibinfo {author} {\bibfnamefont {M.~W.}\ \bibnamefont {Johnson}},
  \bibinfo {author} {\bibfnamefont {S.}~\bibnamefont {Han}}, \bibinfo {author}
  {\bibfnamefont {A.~J.}\ \bibnamefont {Berkley}}, \bibinfo {author}
  {\bibfnamefont {J.}~\bibnamefont {Johansson}}, \bibinfo {author}
  {\bibfnamefont {P.}~\bibnamefont {Bunyk}}, \bibinfo {author} {\bibfnamefont
  {E.}~\bibnamefont {Ladizinsky}}, \bibinfo {author} {\bibfnamefont
  {S.}~\bibnamefont {Govorkov}}, \bibinfo {author} {\bibfnamefont {M.~C.}\
  \bibnamefont {Thom}}, \bibinfo {author} {\bibfnamefont {S.}~\bibnamefont
  {Uchaikin}}, \bibinfo {author} {\bibfnamefont {B.}~\bibnamefont {Bumble}},
  \bibinfo {author} {\bibfnamefont {A.}~\bibnamefont {Fung}}, \bibinfo {author}
  {\bibfnamefont {A.}~\bibnamefont {Kaul}}, \bibinfo {author} {\bibfnamefont
  {A.}~\bibnamefont {Kleinsasser}}, \bibinfo {author} {\bibfnamefont
  {M.~H.~S.}\ \bibnamefont {Amin}},\ and\ \bibinfo {author} {\bibfnamefont
  {D.~V.}\ \bibnamefont {Averin}},\ }\bibfield  {title} {\bibinfo {title}
  {Probing {{Noise}} in {{Flux Qubits}} via {{Macroscopic Resonant
  Tunneling}}},\ }\href {https://doi.org/10.1103/PhysRevLett.101.117003}
  {\bibfield  {journal} {\bibinfo  {journal} {Physical Review Letters}\
  }\textbf {\bibinfo {volume} {101}},\ \bibinfo {pages} {117003} (\bibinfo
  {year} {2008})}\BibitemShut {NoStop}%
\bibitem [{\citenamefont {Kairys}\ \emph {et~al.}(2020)\citenamefont {Kairys},
  \citenamefont {King}, \citenamefont {Ozfidan}, \citenamefont {Boothby},
  \citenamefont {Raymond}, \citenamefont {Banerjee},\ and\ \citenamefont
  {Humble}}]{kairysSimulatingShastrySutherlandIsing2020}%
  \BibitemOpen
  \bibfield  {author} {\bibinfo {author} {\bibfnamefont {P.}~\bibnamefont
  {Kairys}}, \bibinfo {author} {\bibfnamefont {A.~D.}\ \bibnamefont {King}},
  \bibinfo {author} {\bibfnamefont {I.}~\bibnamefont {Ozfidan}}, \bibinfo
  {author} {\bibfnamefont {K.}~\bibnamefont {Boothby}}, \bibinfo {author}
  {\bibfnamefont {J.}~\bibnamefont {Raymond}}, \bibinfo {author} {\bibfnamefont
  {A.}~\bibnamefont {Banerjee}},\ and\ \bibinfo {author} {\bibfnamefont
  {T.~S.}\ \bibnamefont {Humble}},\ }\bibfield  {title} {\bibinfo {title}
  {Simulating the {{Shastry}}-{{Sutherland Ising Model}} using {{Quantum
  Annealing}}},\ }\href {https://doi.org/10.1103/PRXQuantum.1.020320}
  {\bibfield  {journal} {\bibinfo  {journal} {PRX Quantum}\ }\textbf {\bibinfo
  {volume} {1}},\ \bibinfo {pages} {020320} (\bibinfo {year} {2020})},\ \Eprint
  {https://arxiv.org/abs/2003.01019} {arXiv:2003.01019} \BibitemShut {NoStop}%
\bibitem [{\citenamefont {Crowley}\ \emph {et~al.}(2014)\citenamefont
  {Crowley}, \citenamefont {{\DJ}uri{\'c}}, \citenamefont {Vinci},
  \citenamefont {Warburton},\ and\ \citenamefont
  {Green}}]{crowleyQuantumClassicalDynamics2014}%
  \BibitemOpen
  \bibfield  {author} {\bibinfo {author} {\bibfnamefont {P.~J.~D.}\
  \bibnamefont {Crowley}}, \bibinfo {author} {\bibfnamefont {T.}~\bibnamefont
  {{\DJ}uri{\'c}}}, \bibinfo {author} {\bibfnamefont {W.}~\bibnamefont
  {Vinci}}, \bibinfo {author} {\bibfnamefont {P.~A.}\ \bibnamefont
  {Warburton}},\ and\ \bibinfo {author} {\bibfnamefont {A.~G.}\ \bibnamefont
  {Green}},\ }\bibfield  {title} {\bibinfo {title} {Quantum and classical
  dynamics in adiabatic computation},\ }\href
  {https://doi.org/10.1103/PhysRevA.90.042317} {\bibfield  {journal} {\bibinfo
  {journal} {Physical Review A}\ }\textbf {\bibinfo {volume} {90}},\ \bibinfo
  {pages} {042317} (\bibinfo {year} {2014})}\BibitemShut {NoStop}%
\bibitem [{\citenamefont {Albash}\ and\ \citenamefont
  {Lidar}(2015)}]{albashDecoherenceAdiabaticQuantum2015}%
  \BibitemOpen
  \bibfield  {author} {\bibinfo {author} {\bibfnamefont {T.}~\bibnamefont
  {Albash}}\ and\ \bibinfo {author} {\bibfnamefont {D.~A.}\ \bibnamefont
  {Lidar}},\ }\bibfield  {title} {\bibinfo {title} {Decoherence in adiabatic
  quantum computation},\ }\href {https://doi.org/10.1103/PhysRevA.91.062320}
  {\bibfield  {journal} {\bibinfo  {journal} {Physical Review A}\ }\textbf
  {\bibinfo {volume} {91}},\ \bibinfo {pages} {062320} (\bibinfo {year}
  {2015})}\BibitemShut {NoStop}%
\bibitem [{\citenamefont {Zaborniak}\ and\ \citenamefont {{de
  Sousa}}(2021)}]{zaborniakBenchmarkingHamiltonianNoise2021}%
  \BibitemOpen
  \bibfield  {author} {\bibinfo {author} {\bibfnamefont {T.}~\bibnamefont
  {Zaborniak}}\ and\ \bibinfo {author} {\bibfnamefont {R.}~\bibnamefont {{de
  Sousa}}},\ }\bibfield  {title} {\bibinfo {title} {Benchmarking {{Hamiltonian
  Noise}} in the {{D}}-{{Wave Quantum Annealer}}},\ }\href
  {https://doi.org/10.1109/TQE.2021.3050449} {\bibfield  {journal} {\bibinfo
  {journal} {IEEE Transactions on Quantum Engineering}\ }\textbf {\bibinfo
  {volume} {2}},\ \bibinfo {pages} {1} (\bibinfo {year} {2021})}\BibitemShut
  {NoStop}%
\bibitem [{\citenamefont {Lanting}\ \emph {et~al.}(2011)\citenamefont
  {Lanting}, \citenamefont {Amin}, \citenamefont {Johnson}, \citenamefont
  {Altomare}, \citenamefont {Berkley}, \citenamefont {Gildert}, \citenamefont
  {Harris}, \citenamefont {Johansson}, \citenamefont {Bunyk}, \citenamefont
  {Ladizinsky}, \citenamefont {Tolkacheva},\ and\ \citenamefont
  {Averin}}]{lantingProbingHighfrequencyNoise2011}%
  \BibitemOpen
  \bibfield  {author} {\bibinfo {author} {\bibfnamefont {T.}~\bibnamefont
  {Lanting}}, \bibinfo {author} {\bibfnamefont {M.~H.~S.}\ \bibnamefont
  {Amin}}, \bibinfo {author} {\bibfnamefont {M.~W.}\ \bibnamefont {Johnson}},
  \bibinfo {author} {\bibfnamefont {F.}~\bibnamefont {Altomare}}, \bibinfo
  {author} {\bibfnamefont {A.~J.}\ \bibnamefont {Berkley}}, \bibinfo {author}
  {\bibfnamefont {S.}~\bibnamefont {Gildert}}, \bibinfo {author} {\bibfnamefont
  {R.}~\bibnamefont {Harris}}, \bibinfo {author} {\bibfnamefont
  {J.}~\bibnamefont {Johansson}}, \bibinfo {author} {\bibfnamefont
  {P.}~\bibnamefont {Bunyk}}, \bibinfo {author} {\bibfnamefont
  {E.}~\bibnamefont {Ladizinsky}}, \bibinfo {author} {\bibfnamefont
  {E.}~\bibnamefont {Tolkacheva}},\ and\ \bibinfo {author} {\bibfnamefont
  {D.~V.}\ \bibnamefont {Averin}},\ }\bibfield  {title} {\bibinfo {title}
  {Probing high-frequency noise with macroscopic resonant tunneling},\ }\href
  {https://doi.org/10.1103/PhysRevB.83.180502} {\bibfield  {journal} {\bibinfo
  {journal} {Physical Review B}\ }\textbf {\bibinfo {volume} {83}},\ \bibinfo
  {pages} {180502(R)} (\bibinfo {year} {2011})}\BibitemShut {NoStop}%
\bibitem [{\citenamefont {Dickson}\ \emph {et~al.}(2013)\citenamefont
  {Dickson}, \citenamefont {Johnson}, \citenamefont {Amin}, \citenamefont
  {Harris}, \citenamefont {Altomare}, \citenamefont {Berkley}, \citenamefont
  {Bunyk}, \citenamefont {Cai}, \citenamefont {Chapple}, \citenamefont
  {Chavez}, \citenamefont {Cioata}, \citenamefont {Cirip}, \citenamefont
  {{deBuen}}, \citenamefont {{Drew-Brook}}, \citenamefont {Enderud},
  \citenamefont {Gildert}, \citenamefont {Hamze}, \citenamefont {Hilton},
  \citenamefont {Hoskinson}, \citenamefont {Karimi}, \citenamefont
  {Ladizinsky}, \citenamefont {Ladizinsky}, \citenamefont {Lanting},
  \citenamefont {Mahon}, \citenamefont {Neufeld}, \citenamefont {Oh},
  \citenamefont {Perminov}, \citenamefont {Petroff}, \citenamefont {Przybysz},
  \citenamefont {Rich}, \citenamefont {Spear}, \citenamefont {Tcaciuc},
  \citenamefont {Thom}, \citenamefont {Tolkacheva}, \citenamefont {Uchaikin},
  \citenamefont {Wang}, \citenamefont {Wilson}, \citenamefont {Merali},\ and\
  \citenamefont {Rose}}]{dicksonThermallyAssistedQuantum2013}%
  \BibitemOpen
  \bibfield  {author} {\bibinfo {author} {\bibfnamefont {N.~G.}\ \bibnamefont
  {Dickson}}, \bibinfo {author} {\bibfnamefont {M.~W.}\ \bibnamefont
  {Johnson}}, \bibinfo {author} {\bibfnamefont {M.~H.}\ \bibnamefont {Amin}},
  \bibinfo {author} {\bibfnamefont {R.}~\bibnamefont {Harris}}, \bibinfo
  {author} {\bibfnamefont {F.}~\bibnamefont {Altomare}}, \bibinfo {author}
  {\bibfnamefont {A.~J.}\ \bibnamefont {Berkley}}, \bibinfo {author}
  {\bibfnamefont {P.}~\bibnamefont {Bunyk}}, \bibinfo {author} {\bibfnamefont
  {J.}~\bibnamefont {Cai}}, \bibinfo {author} {\bibfnamefont {E.~M.}\
  \bibnamefont {Chapple}}, \bibinfo {author} {\bibfnamefont {P.}~\bibnamefont
  {Chavez}}, \bibinfo {author} {\bibfnamefont {F.}~\bibnamefont {Cioata}},
  \bibinfo {author} {\bibfnamefont {T.}~\bibnamefont {Cirip}}, \bibinfo
  {author} {\bibfnamefont {P.}~\bibnamefont {{deBuen}}}, \bibinfo {author}
  {\bibfnamefont {M.}~\bibnamefont {{Drew-Brook}}}, \bibinfo {author}
  {\bibfnamefont {C.}~\bibnamefont {Enderud}}, \bibinfo {author} {\bibfnamefont
  {S.}~\bibnamefont {Gildert}}, \bibinfo {author} {\bibfnamefont
  {F.}~\bibnamefont {Hamze}}, \bibinfo {author} {\bibfnamefont {J.~P.}\
  \bibnamefont {Hilton}}, \bibinfo {author} {\bibfnamefont {E.}~\bibnamefont
  {Hoskinson}}, \bibinfo {author} {\bibfnamefont {K.}~\bibnamefont {Karimi}},
  \bibinfo {author} {\bibfnamefont {E.}~\bibnamefont {Ladizinsky}}, \bibinfo
  {author} {\bibfnamefont {N.}~\bibnamefont {Ladizinsky}}, \bibinfo {author}
  {\bibfnamefont {T.}~\bibnamefont {Lanting}}, \bibinfo {author} {\bibfnamefont
  {T.}~\bibnamefont {Mahon}}, \bibinfo {author} {\bibfnamefont
  {R.}~\bibnamefont {Neufeld}}, \bibinfo {author} {\bibfnamefont
  {T.}~\bibnamefont {Oh}}, \bibinfo {author} {\bibfnamefont {I.}~\bibnamefont
  {Perminov}}, \bibinfo {author} {\bibfnamefont {C.}~\bibnamefont {Petroff}},
  \bibinfo {author} {\bibfnamefont {A.}~\bibnamefont {Przybysz}}, \bibinfo
  {author} {\bibfnamefont {C.}~\bibnamefont {Rich}}, \bibinfo {author}
  {\bibfnamefont {P.}~\bibnamefont {Spear}}, \bibinfo {author} {\bibfnamefont
  {A.}~\bibnamefont {Tcaciuc}}, \bibinfo {author} {\bibfnamefont {M.~C.}\
  \bibnamefont {Thom}}, \bibinfo {author} {\bibfnamefont {E.}~\bibnamefont
  {Tolkacheva}}, \bibinfo {author} {\bibfnamefont {S.}~\bibnamefont
  {Uchaikin}}, \bibinfo {author} {\bibfnamefont {J.}~\bibnamefont {Wang}},
  \bibinfo {author} {\bibfnamefont {A.~B.}\ \bibnamefont {Wilson}}, \bibinfo
  {author} {\bibfnamefont {Z.}~\bibnamefont {Merali}},\ and\ \bibinfo {author}
  {\bibfnamefont {G.}~\bibnamefont {Rose}},\ }\bibfield  {title} {\bibinfo
  {title} {Thermally assisted quantum annealing of a 16-qubit problem},\ }\href
  {https://doi.org/10.1038/ncomms2920} {\bibfield  {journal} {\bibinfo
  {journal} {Nature Communications}\ }\textbf {\bibinfo {volume} {4}},\
  \bibinfo {pages} {1903} (\bibinfo {year} {2013})}\BibitemShut {NoStop}%
\bibitem [{\citenamefont {Marshall}\ \emph {et~al.}(2019)\citenamefont
  {Marshall}, \citenamefont {Venturelli}, \citenamefont {Hen},\ and\
  \citenamefont {Rieffel}}]{marshallPowerPausingAdvancing2019}%
  \BibitemOpen
  \bibfield  {author} {\bibinfo {author} {\bibfnamefont {J.}~\bibnamefont
  {Marshall}}, \bibinfo {author} {\bibfnamefont {D.}~\bibnamefont
  {Venturelli}}, \bibinfo {author} {\bibfnamefont {I.}~\bibnamefont {Hen}},\
  and\ \bibinfo {author} {\bibfnamefont {E.~G.}\ \bibnamefont {Rieffel}},\
  }\bibfield  {title} {\bibinfo {title} {Power of {{Pausing}}: Advancing
  {{Understanding}} of {{Thermalization}} in {{Experimental Quantum
  Annealers}}},\ }\href {https://doi.org/10.1103/PhysRevApplied.11.044083}
  {\bibfield  {journal} {\bibinfo  {journal} {Physical Review Applied}\
  }\textbf {\bibinfo {volume} {11}},\ \bibinfo {pages} {044083} (\bibinfo
  {year} {2019})}\BibitemShut {NoStop}%
\bibitem [{\citenamefont {Albash}\ and\ \citenamefont
  {Marshall}(2021)}]{albashComparingRelaxationMechanisms2021}%
  \BibitemOpen
  \bibfield  {author} {\bibinfo {author} {\bibfnamefont {T.}~\bibnamefont
  {Albash}}\ and\ \bibinfo {author} {\bibfnamefont {J.}~\bibnamefont
  {Marshall}},\ }\bibfield  {title} {\bibinfo {title} {Comparing relaxation
  mechanisms in quantum and classical transverse-field annealing},\ }\href
  {https://doi.org/10.1103/PhysRevApplied.15.014029} {\bibfield  {journal}
  {\bibinfo  {journal} {Physical Review Applied}\ }\textbf {\bibinfo {volume}
  {15}},\ \bibinfo {pages} {014029} (\bibinfo {year} {2021})},\ \Eprint
  {https://arxiv.org/abs/2009.04934} {arXiv:2009.04934} \BibitemShut {NoStop}%
\bibitem [{\citenamefont {Chen}\ and\ \citenamefont
  {Lidar}(2020{\natexlab{a}})}]{chenWhyWhenPausing2020}%
  \BibitemOpen
  \bibfield  {author} {\bibinfo {author} {\bibfnamefont {H.}~\bibnamefont
  {Chen}}\ and\ \bibinfo {author} {\bibfnamefont {D.~A.}\ \bibnamefont
  {Lidar}},\ }\bibfield  {title} {\bibinfo {title} {Why and when is pausing
  beneficial in quantum annealing?},\ }\href@noop {} {\bibfield  {journal}
  {\bibinfo  {journal} {arXiv:2005.01888 [quant-ph]}\ } (\bibinfo {year}
  {2020}{\natexlab{a}})},\ \Eprint {https://arxiv.org/abs/2005.01888}
  {arXiv:2005.01888 [quant-ph]} \BibitemShut {NoStop}%
\bibitem [{\citenamefont {Chancellor}\ and\ \citenamefont
  {Kendon}(2021)}]{chancellorExperimentalTestSearch2021}%
  \BibitemOpen
  \bibfield  {author} {\bibinfo {author} {\bibfnamefont {N.}~\bibnamefont
  {Chancellor}}\ and\ \bibinfo {author} {\bibfnamefont {V.}~\bibnamefont
  {Kendon}},\ }\bibfield  {title} {\bibinfo {title} {Experimental test of
  search range in quantum annealing},\ }\href
  {https://doi.org/10.1103/PhysRevA.104.012604} {\bibfield  {journal} {\bibinfo
   {journal} {Physical Review A}\ }\textbf {\bibinfo {volume} {104}},\ \bibinfo
  {pages} {012604} (\bibinfo {year} {2021})}\BibitemShut {NoStop}%
\bibitem [{\citenamefont {Shin}\ \emph {et~al.}(2014)\citenamefont {Shin},
  \citenamefont {Smith}, \citenamefont {Smolin},\ and\ \citenamefont
  {Vazirani}}]{shinHowQuantumDWave2014}%
  \BibitemOpen
  \bibfield  {author} {\bibinfo {author} {\bibfnamefont {S.~W.}\ \bibnamefont
  {Shin}}, \bibinfo {author} {\bibfnamefont {G.}~\bibnamefont {Smith}},
  \bibinfo {author} {\bibfnamefont {J.~A.}\ \bibnamefont {Smolin}},\ and\
  \bibinfo {author} {\bibfnamefont {U.}~\bibnamefont {Vazirani}},\ }\bibfield
  {title} {\bibinfo {title} {How "{{Quantum}}" is the {{D}}-{{Wave
  Machine}}?},\ }\href@noop {} {\bibfield  {journal} {\bibinfo  {journal}
  {arXiv:1401.7087 [quant-ph]}\ } (\bibinfo {year} {2014})},\ \Eprint
  {https://arxiv.org/abs/1401.7087} {arXiv:1401.7087 [quant-ph]} \BibitemShut
  {NoStop}%
\bibitem [{\citenamefont {Muthukrishnan}\ \emph {et~al.}(2016)\citenamefont
  {Muthukrishnan}, \citenamefont {Albash},\ and\ \citenamefont
  {Lidar}}]{muthukrishnanTunnelingSpeedupQuantum2016}%
  \BibitemOpen
  \bibfield  {author} {\bibinfo {author} {\bibfnamefont {S.}~\bibnamefont
  {Muthukrishnan}}, \bibinfo {author} {\bibfnamefont {T.}~\bibnamefont
  {Albash}},\ and\ \bibinfo {author} {\bibfnamefont {D.~A.}\ \bibnamefont
  {Lidar}},\ }\bibfield  {title} {\bibinfo {title} {Tunneling and {{Speedup}}
  in {{Quantum Optimization}} for {{Permutation}}-{{Symmetric Problems}}},\
  }\href {https://doi.org/10.1103/PhysRevX.6.031010} {\bibfield  {journal}
  {\bibinfo  {journal} {Physical Review X}\ }\textbf {\bibinfo {volume} {6}},\
  \bibinfo {pages} {031010} (\bibinfo {year} {2016})}\BibitemShut {NoStop}%
\bibitem [{\citenamefont {Albash}\ \emph {et~al.}(2012)\citenamefont {Albash},
  \citenamefont {Boixo}, \citenamefont {Lidar},\ and\ \citenamefont
  {Zanardi}}]{albashQuantumAdiabaticMarkovian2012}%
  \BibitemOpen
  \bibfield  {author} {\bibinfo {author} {\bibfnamefont {T.}~\bibnamefont
  {Albash}}, \bibinfo {author} {\bibfnamefont {S.}~\bibnamefont {Boixo}},
  \bibinfo {author} {\bibfnamefont {D.~A.}\ \bibnamefont {Lidar}},\ and\
  \bibinfo {author} {\bibfnamefont {P.}~\bibnamefont {Zanardi}},\ }\bibfield
  {title} {\bibinfo {title} {Quantum adiabatic {{Markovian}} master
  equations},\ }\href {https://doi.org/10.1088/1367-2630/14/12/123016}
  {\bibfield  {journal} {\bibinfo  {journal} {New Journal of Physics}\ }\textbf
  {\bibinfo {volume} {14}},\ \bibinfo {pages} {123016} (\bibinfo {year}
  {2012})}\BibitemShut {NoStop}%
\bibitem [{\citenamefont {Boixo}\ \emph {et~al.}(2013)\citenamefont {Boixo},
  \citenamefont {Albash}, \citenamefont {Spedalieri}, \citenamefont
  {Chancellor},\ and\ \citenamefont
  {Lidar}}]{boixoExperimentalSignatureProgrammable2013}%
  \BibitemOpen
  \bibfield  {author} {\bibinfo {author} {\bibfnamefont {S.}~\bibnamefont
  {Boixo}}, \bibinfo {author} {\bibfnamefont {T.}~\bibnamefont {Albash}},
  \bibinfo {author} {\bibfnamefont {F.~M.}\ \bibnamefont {Spedalieri}},
  \bibinfo {author} {\bibfnamefont {N.}~\bibnamefont {Chancellor}},\ and\
  \bibinfo {author} {\bibfnamefont {D.~A.}\ \bibnamefont {Lidar}},\ }\bibfield
  {title} {\bibinfo {title} {Experimental signature of programmable quantum
  annealing},\ }\href {https://doi.org/10.1038/ncomms3067} {\bibfield
  {journal} {\bibinfo  {journal} {Nature Communications}\ }\textbf {\bibinfo
  {volume} {4}},\ \bibinfo {pages} {2067} (\bibinfo {year} {2013})}\BibitemShut
  {NoStop}%
\bibitem [{\citenamefont {Albash}\ \emph {et~al.}(2015)\citenamefont {Albash},
  \citenamefont {Vinci}, \citenamefont {Mishra}, \citenamefont {Warburton},\
  and\ \citenamefont {Lidar}}]{albashConsistencyTestsClassical2015}%
  \BibitemOpen
  \bibfield  {author} {\bibinfo {author} {\bibfnamefont {T.}~\bibnamefont
  {Albash}}, \bibinfo {author} {\bibfnamefont {W.}~\bibnamefont {Vinci}},
  \bibinfo {author} {\bibfnamefont {A.}~\bibnamefont {Mishra}}, \bibinfo
  {author} {\bibfnamefont {P.~A.}\ \bibnamefont {Warburton}},\ and\ \bibinfo
  {author} {\bibfnamefont {D.~A.}\ \bibnamefont {Lidar}},\ }\bibfield  {title}
  {\bibinfo {title} {Consistency {{Tests}} of {{Classical}} and {{Quantum
  Models}} for a {{Quantum Annealer}}},\ }\href
  {https://doi.org/10.1103/PhysRevA.91.042314} {\bibfield  {journal} {\bibinfo
  {journal} {Physical Review A}\ }\textbf {\bibinfo {volume} {91}},\ \bibinfo
  {pages} {042314} (\bibinfo {year} {2015})},\ \Eprint
  {https://arxiv.org/abs/1403.4228} {arXiv:1403.4228} \BibitemShut {NoStop}%
\bibitem [{\citenamefont {Kramers}\ and\ \citenamefont
  {Wannier}(1941)}]{kramersStatisticsTwoDimensionalFerromagnet1941}%
  \BibitemOpen
  \bibfield  {author} {\bibinfo {author} {\bibfnamefont {H.~A.}\ \bibnamefont
  {Kramers}}\ and\ \bibinfo {author} {\bibfnamefont {G.~H.}\ \bibnamefont
  {Wannier}},\ }\bibfield  {title} {\bibinfo {title} {Statistics of the
  {{Two}}-{{Dimensional Ferromagnet}}. {{Part I}}},\ }\href
  {https://doi.org/10.1103/PhysRev.60.252} {\bibfield  {journal} {\bibinfo
  {journal} {Physical Review}\ }\textbf {\bibinfo {volume} {60}},\ \bibinfo
  {pages} {252} (\bibinfo {year} {1941})}\BibitemShut {NoStop}%
\bibitem [{\citenamefont
  {Klauder}(1979)}]{klauderPathIntegralsStationaryphase1979}%
  \BibitemOpen
  \bibfield  {author} {\bibinfo {author} {\bibfnamefont {J.~R.}\ \bibnamefont
  {Klauder}},\ }\bibfield  {title} {\bibinfo {title} {Path integrals and
  stationary-phase approximations},\ }\href
  {https://doi.org/10.1103/PhysRevD.19.2349} {\bibfield  {journal} {\bibinfo
  {journal} {Physical Review D}\ }\textbf {\bibinfo {volume} {19}},\ \bibinfo
  {pages} {2349} (\bibinfo {year} {1979})}\BibitemShut {NoStop}%
\bibitem [{\citenamefont {Kato}(1950)}]{katoAdiabaticTheoremQuantum1950}%
  \BibitemOpen
  \bibfield  {author} {\bibinfo {author} {\bibfnamefont {T.}~\bibnamefont
  {Kato}},\ }\bibfield  {title} {\bibinfo {title} {On the {{Adiabatic Theorem}}
  of {{Quantum Mechanics}}},\ }\href {https://doi.org/10.1143/JPSJ.5.435}
  {\bibfield  {journal} {\bibinfo  {journal} {Journal of the Physical Society
  of Japan}\ }\textbf {\bibinfo {volume} {5}},\ \bibinfo {pages} {435}
  (\bibinfo {year} {1950})}\BibitemShut {NoStop}%
\bibitem [{\citenamefont {Chen}\ and\ \citenamefont
  {Lidar}(2020{\natexlab{b}})}]{chenHOQSTHamiltonianOpen2020}%
  \BibitemOpen
  \bibfield  {author} {\bibinfo {author} {\bibfnamefont {H.}~\bibnamefont
  {Chen}}\ and\ \bibinfo {author} {\bibfnamefont {D.~A.}\ \bibnamefont
  {Lidar}},\ }\bibfield  {title} {\bibinfo {title} {{{HOQST}}: Hamiltonian
  {{Open Quantum System Toolkit}}},\ }\href@noop {} {\bibfield  {journal}
  {\bibinfo  {journal} {arXiv:2011.14046 [quant-ph]}\ } (\bibinfo {year}
  {2020}{\natexlab{b}})},\ \Eprint {https://arxiv.org/abs/2011.14046}
  {arXiv:2011.14046 [quant-ph]} \BibitemShut {NoStop}%
\bibitem [{\citenamefont {Mishra}\ \emph {et~al.}(2018)\citenamefont {Mishra},
  \citenamefont {Albash},\ and\ \citenamefont
  {Lidar}}]{mishraFiniteTemperatureQuantum2018}%
  \BibitemOpen
  \bibfield  {author} {\bibinfo {author} {\bibfnamefont {A.}~\bibnamefont
  {Mishra}}, \bibinfo {author} {\bibfnamefont {T.}~\bibnamefont {Albash}},\
  and\ \bibinfo {author} {\bibfnamefont {D.~A.}\ \bibnamefont {Lidar}},\
  }\bibfield  {title} {\bibinfo {title} {Finite temperature quantum annealing
  solving exponentially small gap problem with non-monotonic success
  probability},\ }\href {https://doi.org/10.1038/s41467-018-05239-9} {\bibfield
   {journal} {\bibinfo  {journal} {Nature Communications}\ }\textbf {\bibinfo
  {volume} {9}},\ \bibinfo {pages} {2917} (\bibinfo {year} {2018})}\BibitemShut
  {NoStop}%
\bibitem [{\citenamefont {Zhou}\ \emph {et~al.}(2020)\citenamefont {Zhou},
  \citenamefont {Wang}, \citenamefont {Choi}, \citenamefont {Pichler},\ and\
  \citenamefont {Lukin}}]{zhouQuantumApproximateOptimization2020}%
  \BibitemOpen
  \bibfield  {author} {\bibinfo {author} {\bibfnamefont {L.}~\bibnamefont
  {Zhou}}, \bibinfo {author} {\bibfnamefont {S.-T.}\ \bibnamefont {Wang}},
  \bibinfo {author} {\bibfnamefont {S.}~\bibnamefont {Choi}}, \bibinfo {author}
  {\bibfnamefont {H.}~\bibnamefont {Pichler}},\ and\ \bibinfo {author}
  {\bibfnamefont {M.~D.}\ \bibnamefont {Lukin}},\ }\bibfield  {title} {\bibinfo
  {title} {Quantum {{Approximate Optimization Algorithm}}: Performance,
  {{Mechanism}}, and {{Implementation}} on {{Near}}-{{Term Devices}}},\ }\href
  {https://doi.org/10.1103/PhysRevX.10.021067} {\bibfield  {journal} {\bibinfo
  {journal} {Physical Review X}\ }\textbf {\bibinfo {volume} {10}},\ \bibinfo
  {pages} {021067} (\bibinfo {year} {2020})}\BibitemShut {NoStop}%
\end{thebibliography}%

\appendix

\section{PFC Free-Energy Proof}
\label{Appendix_free_energy}

The transfer matrix representation of the PFC partition function (Eq.~(\ref{eq:classical_Z})) involves singular, non-commuting matrices $\mathbf{W}$ (Eq.~(\ref{eq: transfer_matrix})) and $\mathbf{V} = \mathbf{v}^T \mathbf{v}$ (where $\mathbf{v}$ is defined in Eq.~(\ref{eq: boundary_vec})), which does not allow for an obvious reduction to an analytical free energy that is generally defined as
\begin{equation}
    F = -\lim_{N \rightarrow \infty} \frac{1}{\beta N} \ln \mathcal{Z}\,,
    \label{eq: free_energy_app}
\end{equation}
where $\mathcal{Z}$ is the partition function. We begin by redefining the partition function in Eq.~(\ref{eq:classical_Z}) to
\begin{equation}
    \mathcal{Z} = \textrm{Tr}\left(\mathbf{W}^{M-1} \mathbf{V}\right)\,,
\end{equation}
where $M = N/2$, such that we now take the limit in $M \rightarrow \infty$ due to $\mathbf{W}$ containing information about the subsystem rather than a single qubit. Performing an eigen-decomposition on $\mathbf{W}$ to yields a diagonal matrix of eigenvalues, $\mathbf{D}$, and a matrix of eigenvectors, $\mathbf{P}$, in the form $\mathbf{W} = \mathbf{P}\mathbf{D}\mathbf{P}^{-1}$. 
\begin{figure}
    \centering
    \includegraphics[width=1\columnwidth]{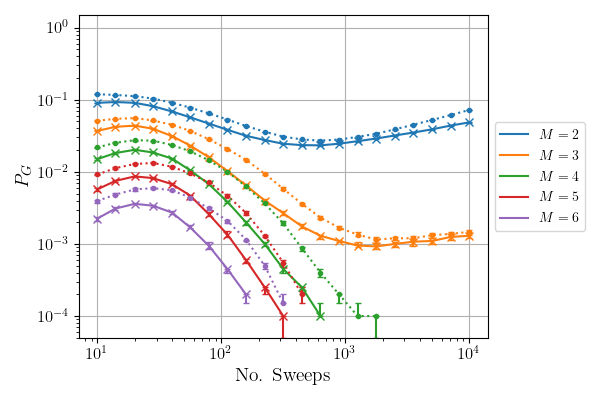}
    \caption{Probability of being in the ground state for both SVMC-TF (solid line) and spherical-SVMC (dotted line) as the system scales in size, $M$, for a PFC with $d = 0.1$. Here, the SVMC and spherical-SVMC-TF probabilities are found from 20,000 samples, which we repeat 50 times and bootstrap to find the median and $95\%$ confidence intervals for the data point and error bars respectively.}
    \label{fig:result_svmc_n_test2}
\end{figure}
\begin{figure*}
    \centering
    \includegraphics[width=2\columnwidth]{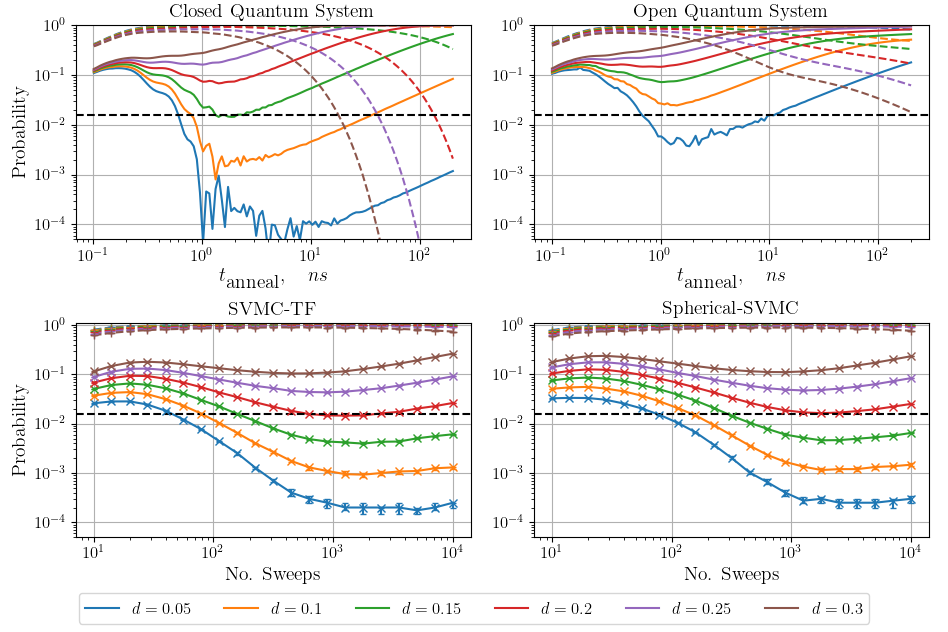}
    \caption{Plots of state probability for being in either the ground state (solid lines) or any of the $2^M$-degenerate $1^{st}$ excited states (dashed lines) at the end of an anneal. A PFC with system size $M = 3$  was evolved using quantum (top row, but now on a log-scale unlike Fig.~\ref{fig:result_quantum_svmc_n=3}) and classical (bottom row) dynamics with SVMC-TF and spherical-SVMC. The black dashed line indicates the probability with random guessing  i.e. $1/64$. The quantum simulations are plotted against anneal time in $ns$, whilst the SVMC simulations are plotted against the number of sweeps. The closed and open system dynamics are evolved according to the von-Neumann and adiabatic master equation respectively (see section~\ref{Methods_analytical}). The SVMC-TF and spherical-SVMC probabilities are found from 20,000 samples, of which we repeat 50 times and bootstrap to find a median and $ 95 \% $ confidence intervals for the data point and error bars respectively.}
    \label{fig:result2_quantum_svmc_n=3}
\end{figure*}
When $\mathbf{W}$ is raised to any power, the decomposition simply becomes $\mathbf{W}^M = \mathbf{P}\mathbf{D}^M\mathbf{P}^{-1}$. Given the cyclic invariance of the trace, the partition function therefore becomes
\begin{equation}
    \mathcal{Z} = \textrm{Tr}\left(\mathbf{P}\mathbf{D}^{M-1}\mathbf{P}^{-1} \mathbf{V}\right) = \textrm{Tr}\left(\mathbf{D}^{M-1}\mathbf{P}^{-1} \mathbf{V} \mathbf{P}\right)\,.
\end{equation}
Taking $\mathbf{A} = \mathbf{P}^{-1} \mathbf{V} \mathbf{P}$, and the largest absolute eigenvalue of $\mathbf{W}$ to be $\lambda_1$ (the spectral radius), the trace summation will yield
\begin{equation}
    \begin{split}
        \textrm{Tr}\left(\mathbf{D}^{M-1}\mathbf{A} \right) = \sum_{i=1}^{4}A_{ii}\lambda_i^{M-1} = \sum_{i=1}^{4}\left(\sqrt[\leftroot{-3}\uproot{3} M-1]{A_{ii}}\lambda_i\right)^{M-1} \\
        = \left(\sqrt[\leftroot{-3}\uproot{3} M-1]{A_{11}}\lambda_1\right)^{M-1} \sum_{i=1}^{4}\left(\frac{\sqrt[\leftroot{-3}\uproot{3} M-1]{A_{ii}}\lambda_i}{\sqrt[\leftroot{-3}\uproot{3} M-1]{A_{11}}\lambda_1}\right)^{M-1}\,.
    \end{split}
\end{equation}
Given that $\mathbf{W}$ is positive semi-definite, in the limit of $M \rightarrow \infty$ the trace is simply left with $\lambda_1^{M-1}$ as a non-vanishing term, such that we find the analytical form of the free energy to be
\begin{equation}
    F = -\lim_{M \rightarrow \infty} \frac{1}{\beta M} \ln \textrm{Tr}\left(\mathbf{D}^{M-1}\mathbf{A} \right) = -\frac{1}{\beta} \ln{\lambda_1}\,.
    \label{eq: free_energy_app2}
\end{equation}
This eigenvalue can be found symbolically using Pythons \textit{SymPy} library, such that 
\begin{equation}
\begin{split}
    \lambda_1 =& e^{2\beta R} \cosh{\beta Rd} + \cosh{\beta R(2-d)} +\\ & \sqrt{\left( e^{2\beta R} \cosh{\beta Rd} + \cosh{\beta R(2-d)} \right)^2 - 4\sinh{4\beta R}}\,.
\end{split}
    \label{eq: free_energy_app3}
\end{equation}
The validity of this free-energy relies on the fact that our spectral radius is $\geq 1$, otherwise $\mathbf{W}$ would converge to zero in the limit of $M \rightarrow \infty$. However, the PFC has bounds of $R > 0$ and $0< d < 1$, such that our eigenvalue $\lambda_1 \geq 4$, and is therefore finite everywhere except for $\beta R = \infty$.

\section{Result of SVMC Variants}
\label{Appendix_svmc_var}

In this appendix, we present three additional results of the two other SVMC variants (SVMC-TF and spherical-SVMC) and demonstrate the intermediate effects in performance when going from the simplest case (SVMC) to the most complex (spherical-SVMC-TF) variant. The first is of Fig.~\ref{fig:result_svmc_n_test2}, where we illustrate how SVMC-TF and spherical-SVMC are affected by increasing the system size, similarly to what is seen in Fig.~\ref{fig:result_svmc_n_test}. Secondly, we see the results for SVMC-TF and spherical-SVMC (Fig.~\ref{fig:result2_quantum_svmc_n=3}) missing from our comparative results in Fig.~\ref{fig:result_quantum_svmc_n=3}. This also includes a log-scale view of the quantum systems tested for comparison, which highlights what appears to be discontinuities in the results. These are caused by under-sampling the rapid oscillatory behaviour in the ground state probability after passing through the minimum gap too quickly (given that sufficient probability density is placed on the ground state), and this is amplified for small $d$ (i.e. small minimum gap).

\begin{figure*}
    \centering
    \includegraphics[width=2\columnwidth]{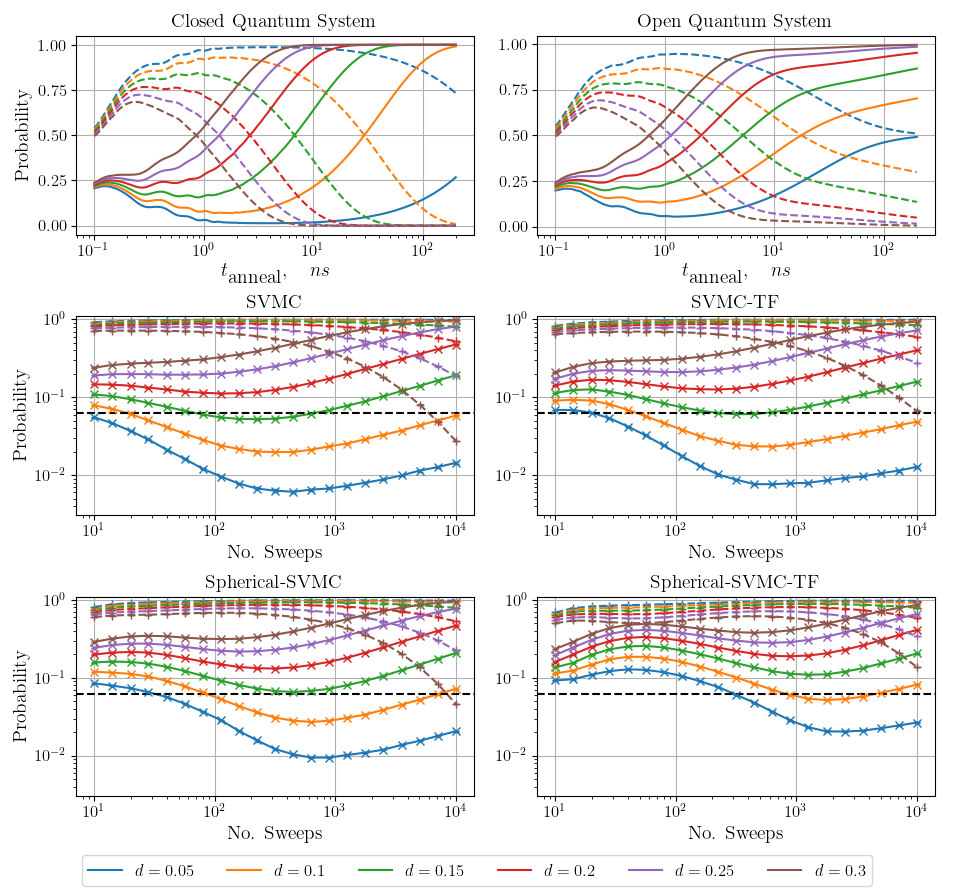}
    \caption{Plots of state probability for being in either the ground state (solid lines) or any of the $1^{st}$ excited states (dashed lines) for the closed and open system quantum simulations  and the SVMC variants for a PFC with system size $M = 2$. The black dashed line indicates the probability with random guessing  i.e. $1/16$. The quantum simulations are measured against anneal time in $ns$, whilst the SVMC simulations are measured against the number of sweeps used to increment the anneal. Here, the probabilities of the SVMC variants are found from 20,000 samples, of which we repeat 50 times and bootstrap to find a median and $95\%$ confidence intervals for the data point and error bars respectively.}
    \label{fig:result_quantum_svmc_n=2}
\end{figure*}

 Finally, we show a comparative experiment for the simpler $M=2$ PFC in Fig.~\ref{fig:result_quantum_svmc_n=2} that is analogous to the $M=3$ case shown in Fig.\ref{fig:result_quantum_svmc_n=3}. Here we tune the perturbative parameter $d$ and show results and simulations for all variants. As expected, the $M=2$ is easier to solve and we therefore see higher ground state probabilities in general compared to the $M=3 $ case. Furthermore, it illustrates better how combining both the transverse-field updates and the extension to the whole Bloch sphere has a enhanced effect and out-performs all variants for the hardest cases (given the sweep ranges we have tested).

From the variants tested, each individual addition to the SVMC algorithm brings an improvement to ground state probability for the hardest problem cases. The helpful addition of the azimuthal component to SVMC can be explained by the fact that not only does it provide another degree of freedom, but on average it can reduce the transverse field energy contribution in Eq.~\ref{eq:SVMC_sph_energy_fn}. This reduces both the size of the minimum gap and moves the position of the minimum gap to earlier in the anneal, therefore giving SVMC more time (sweeps) to try and reach the ground state after the minimum gap. 
 
Finally, we would expect the addition of transverse-field updates to provide a more accurate description of the a quantum annealer (by emulating freeze-out~\cite{albashComparingRelaxationMechanisms2021}), but also hinder computation as dynamics would be slow in regimes where we would need additional dynamics to reach the ground state after the minimum gap. However, provided that the spin-vector has followed the false minimum and is near the global minimum of the semi-classical potential (Fig.~\ref{fig:semi-classical}), the spin-vector would most likely resemble the first excited state closest in Hamming distance to the ground state after the minimum gap (see Fig.~\ref{fig:result_populations}). Therefore, slowed dynamics are preventing the spin-vector from getting lost in the first-excited-state manifold to some extent, which makes it statistically more likely for SVMC-TF to reach the ground state.

\end{document}